\documentclass[aps,showpacs,preprintnumbers,amsmath,amssymb,twocolumn,tightenlines
,eqsecnum
]{revtex4}

\usepackage{bm}
\usepackage{amsmath}
\usepackage{graphicx}
\usepackage{epic}
\usepackage{eepic}

\newcommand{\E}{\mathcal{E}}
\newcommand{\M}{M_{p_+p_-}}
\newcommand{\kp}[2]{\kappa^{#1}_{#2}}
\newcommand{\B}{\mathcal{H}}
\newcommand{\SA}{\mathcal{S}}
\newcommand{\F}{\mathcal{F}}
\newcommand{\e}{\varepsilon}
\newcommand{\p}{\tilde{\bm{p}}}
\newcommand{\G}{\mathcal{G}}

\begin{document}

    \bibliographystyle{apsrev}

    \title {Electron-positron pair creation by Coulomb and laser fields in the tunneling regime}
                
    \author{M.Yu.Kuchiev}
                \email[email:]{kmy@phys.unsw.edu.au}
    \author{D.J.Robinson}
                \email[email:]{robinson@phys.unsw.edu.au}
    \affiliation{School of Physics, University of New South Wales,
      Sydney 2052, Australia}
    \date{\today}

\begin{abstract}
Electron-positron pair creation due to combined nuclear Coulomb and strong laser fields is investigated for the tunneling regime. The energy spectra and angular distributions of the pair are found analytically for the first time. The energy spectrum for each lepton exhibits a sharp maximum located well above the threshold for any polarization of the laser field. The angular distributions of leptons depend on the polarization: for the linear polarization both leptons move predominantly along the laser beam direction; for the circular polarization leptons are emitted in a thin-walled cone centered on the laser beam. The spectral and angular distributions found are governed by the intensity and frequency of the field, and the frequency independent total pair creation rates comply with the previously known results. A new method of calculation - the \emph{vicinal} approximation - which uses the fact that the pair production takes place in the close vicinity of the nucleus, is suggested.
\end{abstract}


    \maketitle
 
\section{Introduction}
\label{sec:I}

We consider the electron-positron pair creation due to a strong laser field and the Coulomb field of a nucleus. This combination of fields may create electron-positron pairs via a multiphoton process, sometimes called the nonlinear Bethe-Heitler process \cite{BH34}. Such a scheme of electron-positron pair creation has received attention over the last decade due to advances in laser technology, which may permit it to be tested in an x-ray free electron laser (XFEL) experiment \cite{Rin01}.

The multiphoton pair creation by combined strong laser and Coulomb fields was first treated by Yakovlev \cite{Yak66}. He considered total and partial cross sections of the pair creation process in the case of a circularly polarized laser field. Mittleman \cite{Mit87} later calculated the cross-section and pair creation rate for the low intensity linear polarization case. More recently, Refs. \cite{MVG03, MVG032, MVG04} have considered the closely related problem of multiphoton pair creation due to a ultra-relativistic nucleus colliding with a strong, circularly polarized laser field. The energy spectra and angular distributions of the produced leptons were calculated numerically over a range of Lorentz factors of the nucleus. Ref. \cite{MVG04} also treated the linear polarization case and the scenario in which the pair is created in a bound state of the nucleus, and Ref. \cite{DP98} considered pair creation in the vicinity of heavy ions. Ref. \cite{AAMS03} has derived the total creation rate due to a circularly polarized laser field incident on a nucleus and also calculated numerically the lepton energy spectrum for strong fields.

One approach widely used in the literature treats the interaction of leptons with the laser field exactly by use of the Volkov states \cite{Vol35,Vol37}. It is analogous to the Keldysh approximation used for the problem of multiphoton ionisation of atoms or ions \cite{Kel65,NR66,PPT66,Fai73,Rei80,GK97,GK972,AK85}.
Alternatively, Milstein \emph{et al.} \cite{MMHJK06} have recently derived pair creation rates and cross-sections for different cases of polarizations and different parameters of the laser field by application of the optical theorem and use of the polarization operator of a photon in a laser field. 

Overall, the total rates of the pair creation in the problem considered have been studied in detail by different methods, which all result in clear, simple analytical expressions. However, the more detailed characteristics of the problem, namely the spectral and angular distributions of the leptons, have not been examined with the same level of detail. Meanwhile, considering future experimental studies of pair creation, which is one of the possible purposes of the XFEL facilities, will require thorough knowledge of these characteristics and their dependence on the parameters of the laser field. 

In the present work, we address the problem of the pair creation in the tunneling regime. We propose a new variant of the Keldysh approach, the so called vicinal aprroximation, which takes into account important physical properties of the system and simultaneously greatly simplifies the calculations required. We then find a complete analytical solution to this problem, deriving new analytical expressions for the angular distributions and energy spectra of the created leptons, as well as photon absorption spectra. The linear and circular laser polarizations are studied in detail, and a brief outline of the general elliptic case is presented. Our preliminary results were discussed in Ref. \cite{Dean}.

\section{Theoretical Framework}
\subsection{Laser Parameters}
\label{sec:TFLP}

The laser frequency $\omega$ is assumed to be small, so that it satisfies the
adiabatic condition
\begin{equation}
        \label{eqn:AR}
        \omega \ll m,
\end{equation}
where $m$ the electron mass. Unless stated otherwise we use relativistic units  $\hbar = c = 1$, $e^2=\alpha, e>0$, where $-e$ is the electron charge. Secondly, the laser electric field strength $\E$ is considered to be small in comparison with the QED critical field $\E_c =m^2/e\equiv 1.3\times10^{18}$ Vm$^{-1}$, so that their ratio
\begin{equation}
        \label{eqn:WF}
         \F \equiv \E/\E_c \ll 1.
\end{equation}
Thirdly, the  adiabatic (intensity) parameter of the problem $\xi$ is presumed to be large:
\begin{equation}
        \label{eqn:IP}
        \xi \equiv  \frac{e\E}{m\omega} \gg 1.
\end{equation}
Eqs.(\ref{eqn:AR})-(\ref{eqn:IP}) together describe the tunneling regime for the pair production. This regime may be achievable in future experiments, since it is expected \cite{CP98,Rin01} that XFEL facilities will be able to produce electric field strengths up to $\E \simeq 10^{-1}\E_c$ at frequency $\omega \simeq 0.002m - 0.02 m$. Conditions (\ref{eqn:AR})-(\ref{eqn:IP}) may be equivalently expressed by the combined inequality
\begin{equation}
        \label{eqn:TR}
        1 \ll \xi \ll m/\omega.
\end{equation}

\subsection
{Amplitude and Probability}
\label{sec:TFCRPA}

Consider the field produced by a plane electromagnetic (EM) wave and the Coulomb field of a bare nucleus in the nuclear rest frame, with the nucleus placed at the coordinate system origin. The nuclear Coulomb potential for an electron at radius $r$ is thus $-Z\alpha/r$, where $Z$ is the atomic number of the nucleus.

The amplitude of a multiphoton process is given by a matrix element averaged over the period  $T = 2\pi/\omega$ of oscillations of the EM plane wave, see e.g. Ref. \cite{GK97,AK85}.
Specifically for the case of the pair creation the amplitude reads  
\begin{equation}
        \label{eqn:PAGF}
        \M = \frac{1}{T}\int_0^T \langle \psi_{\bm{p}_-}(t)\left| -Z\alpha/r\ \right|\psi^{*}_{\bm{p}_+}(t) \rangle\ dt~.
\end{equation}
Here $\psi_{\bm{p}_-}$ and $\psi_{\bm{p}_+}$ are the wavefunctions of the electron and positron.  The subscripts `$-$' and `$+$' henceforth denote electron and positron variables respectively, and the superscript $^*$ denotes a complex conjugate. Note that the electron and positron states are indexed by their momenta $\bm{p}_{\pm}$. The Floquet theorem guarantees that we may always write the wavefunction $\psi_{\bm{p}_\pm}$ of these particles in the form
\begin{equation}
        \label{eqn:PSI}
        \psi_{\bm{p}_\pm}(\bm{r},t) = \varphi_{\bm{p}_{\pm}}(\bm{r},t)\exp\left(-i\bar{E}_{\pm}t\right)~,
\end{equation}
where  $\bar{E}_{\pm}$ is the mean quasienergy for the particle (see Section \ref{sec:TFVA}), and  $\varphi_{\bm{p}_{\pm}}(\bm{r},t)$ is a periodic function of $t$ with the period $T$.

The matrix element $\langle|\,|\rangle$ in the integrand in Eq.(\ref{eqn:PAGF}) means a conventional integral over spatial coordinates, that is
\begin{equation}
        \label{eqn:MED}
        \langle \psi_{\bm{p}_-}(t)\left|1/r \right|\psi^{*}_{\bm{p}_+}(t) \rangle  =\int\psi_{\bm{p}_-}^{*}(\bm{r},t)\psi^{*}_{\bm{p}_+}(\bm{r},t) \frac{d^3r}{r}~.
\end{equation}
Both the electron and positron states are described here by complex-conjugated wave functions since they both represent the created particles.
Together with the momenta $\mathbf{p}_\pm$, each wave function here is characterized also by the spinor index $\lambda_\pm$, which is not depicted explicitly.

The probability of the multiphoton process is proportional to the square of the amplitude, $|\M|^2$ which should be multiplied by appropriate, conventional statistical factors, see e.g. \cite{AK85,GK97} for details. Using this rule one writes the rate of the pair creation, $W$, as
\begin{align}
        \label{eqn:TPCR}
        W
        &= 2\pi \sum_{s_\pm}\,\,\sum_{n=1}^\infty \int \delta(\bar{E}_- + \bar{E}_+ - n\omega)\notag\\
        & \quad \times \sum_{\lambda_\pm}\,|\M|^2 \frac{d^3\bm{p}_-}{(2\pi)^3}\frac{d^3\bm{p}_+}{(2\pi)^3}~.
\end{align}
Here $n$ is the number of the photons absorbed, summation over $s_\pm$ takes into account two spin states of each lepton $s_-,s_+=\pm 1/2$, while summation over $\lambda_\pm$ refers to the spinor indexes (for the sake of simplicity
this summation is not presented in the following formulas explicitly). The Dirac delta-function imposes the energy conservation law
\begin{equation}
        \label{eqn:ECL}
        \bar{E}_+ + \bar{E}_- = n\omega~.
\end{equation}
In the case $\omega \ll m$, Eq. (\ref{eqn:TPCR}) may be simplified by noting that the sum over $n$ can be replaced by a corresponding integral, which gives the creation rate the asymptotic form
\begin{equation}
        \label{eqn:TPCRGF}
        {W} = \frac{2\pi}{\omega}\sum_{s_\pm}\int|\M|^2\frac{d^3\bm{p}_-}{(2\pi)^3}\frac{d^3\bm{p}_+}{(2\pi)^3}~.
\end{equation}
Taking partial summations and integrations here one finds expressions for the differential probabilities $d {W}$, which describe the energy spectra and angular distributions for the pair, as well as the photon absorption spectra.

\subsection{Volkov Wave Functions}
Presuming that the nuclear charge is not large, i.e.
\begin{equation}
        Z\alpha \ll 1~,
\end{equation}
we treat the interaction of leptons with the Coulomb field perturbatively to first order. The lepton state $\psi_{\bm{p}_\pm}$ in this approach, which stems from Refs.\cite{Kel65, Yak66}, can be described by a Volkov wavefunction that takes into account the lepton interaction only with the EM plane wave.

Let $ A = A(\Phi) $ be the 4-vector potential of the electromagnetic plane wave, where the scalar $\Phi = (kx)$, $k$ being here the light 4-wavevector, $k^2=0$, and $x =(t,x,y,z)$ is the usual spacetime coordinate. We use the conventional notation that $a^{\mu}b_{\mu} = (ab)$, and $a^{\mu}a_{\mu}=a^2$. In the Lorentz gauge, $(\partial A) = 0$, the state of the electron in the EM plane wave is described by the Volkov wavefunction $\Psi_-$ (see e.g. Ref. \cite{LL82} for a derivation):
\begin{equation}
        \label{eqn:RVS}
        \Psi_- = \left(1 - \frac{e}{2\ (kp)}(\gamma k)(\gamma A)\right)\exp\left( iS_- \right)
        \frac { u_{\bm{p}}  } {\sqrt{2\e} }~,
\end{equation}
where $e>0$; $p = (\e, \bm{p})$ is the electron 4-momentum (for clarity we temporarily drop the `$\pm$' subscripts for it), $\bm{p} = \big(p_x,p_y,p_z\big)$; $\gamma = (\gamma^{0},\bm{\gamma})$ are the Dirac matrices; $u_{\bm{p}}$ is a free Dirac 4-spinor, which satisfies
\begin{equation}
        (\gamma p)\,u_{\bm{p}}=mu_{\bm{p}}~;
        \label{eqn:DE}
\end{equation}
and $S_-$ is the classical action for the electron in the electromagnetic wave
\begin{equation}
        \label{eqn:RVSVA}
        S_- = -(px) + \int^{(kx)}\frac{e}{(kp)}\left((pA) + \frac{e}{2} A^2\right)d\Phi.
\end{equation}
Henceforth, we assume that the electromagnetic wave propagates along the $z$ direction. Then $k = (\omega,0,0,\omega)$, so that $(kp) = \omega(\e - p_z)$, $\Phi = \omega(t-z)$, and
it is convenient to define
\begin{equation}
        \label{eqn:ED}
        \eta\equiv\e - p_z.
\end{equation}
Equation (\ref{eqn:DE}) implies that $p^2=m^2$, so that the leptons are created on the mass shell. It follows that $\eta > 0$ and
\begin{equation}
        \label{eqn:PZE}
        p_z = \frac{m^2 + \kappa^2-\eta^2}{2\eta},\quad
        \e  = \frac{m^2 + \kappa^2+\eta^2}{2\eta},
\end{equation}
where $\kappa^2 = p_x^2 + p_y^2$.
The 4-vector potential of the electromagnetic wave may be written as
\begin{equation}
        A(\Phi) = \frac{\E}{\omega}\,a(\Phi)=\frac{\E}{\omega}\,\Big(0,\bm{a}(\Phi)\Big),
\end{equation}
where $ \bm{a} (\Phi)$ is the polarization of the wave and we have chosen $A^0 = 0$. This polarization satisfies
$ \bm{a} (\Phi) \cdot \bm{k}=0 $. We may now write the electron wavefunction (\ref{eqn:RVS}) and action (\ref{eqn:RVSVA})
in a clearer form:
\begin{align}
        \Psi_- &= \left(1 - \frac{m\,\xi}{2\eta}(\gamma^0 -\gamma^3)(\gamma a)\right)\exp(iS_-)\frac{u_{\bm{p}} }{\sqrt{2\e}}~,\label{eqn:RVSP}\\
        S_- & = -(px)  + \frac{m\xi}{\omega\eta}\int^{\omega(t-z)}\!\!\bigg((pa) + \frac{m\xi}{2}a^2\bigg)d\Phi~.         \label{eqn:RVSPVA}
\end{align}
Similarly, the positron wave function and the action read
\begin{align}
        \Psi_{+}
        &= \bigg(\!1 + \frac{m\,\xi}{2\eta}(\gamma^0 -\gamma^3)
        (\gamma a) \bigg) \exp(iS_+)
        \frac{  \tilde{u}_{\bm{p}}}
        {\sqrt{2\e}}~,\label{eqn:posWF}\\
        S_+ & = -(px)  - \frac{m\xi}{\omega\eta}\int^{\omega(t-z)}
        \bigg((pa) - \frac{m\xi}{2}a^2\bigg)d\Phi~. \label{eqn:posS}
\end{align}
The spinor $\tilde{u}_{\bm{p}}$ satisfies the Dirac equation (\ref{eqn:DE}) describing the propagation of a free positron, but in calculations 
it is convenient to express it conventionlly, via the spinor that describes an electron in the "lower" continuum using an operator of the charge conjugation $\hat {C}\psi =\gamma^2\psi^*$,
\begin{equation}
\label{C}
\tilde{u}_{\bm{p}}( \varepsilon_{\bm{p}})=\gamma^2
{u}_{-\bm{p}}^*( -\varepsilon_{\bm{p}})~.
\end{equation}
Here we have to specify an energy $\pm \varepsilon_{\bm{p}}=\pm(\bm{p}^2+m^2)^{1/2}$ of the state described by the corresponding spinor; the spinor ${u}_{-\bm{p}}^*( -\varepsilon_{\bm{p}})$ describes an electron state with negative energy, i. e. the state in the lower continuum. (For clarity spinors are presented below without reference to the energy of the state they describe.)

\subsection{Vicinal approximation}
\label{sec:TFVA}
The interaction of leptons with the Coulomb field must be strong, for otherwise the probability of the pair creation would be very small. We therefore presume that the most important events during the pair creation take place in the close vicinity of the nucleus, i.e. we presume that the pair is created at distances which are comparable or smaller than the Compton radius  $r_c = 1/m$. We will verify later that this presumption is correct by direct calculation (see Eqs. \ref{eqn:LPQMSP} and \ref{eqn:QMSP}).

Having this presumption in mind we can presume now that $|z| \le 1/m$ in
Eqs. (\ref{eqn:RVSP}) and (\ref{eqn:RVSPVA}). Importantly, in the adiabatic regime $\omega \ll m$, so it follows that
\begin{equation}
        \label{eqn:VA}
        \omega |z| \ll 1.
\end{equation}
We call inequality (\ref{eqn:VA}) the \emph{vicinal} approximation, since it arises from the close vicinity of pair creation to the nucleus.

Since $\omega z $ is small, we can expand the action (\ref{eqn:RVSPVA}) in a Taylor series about $\Phi = \omega t$ to first order in $\omega z$, and replace argument $\omega (t-z)$ with $\omega t$ in the pre-exponential term of the wave function (\ref{eqn:RVSP}).  Thus in the vicinal approximation the Volkov wavefunction of the electron is greatly simplified:
\begin{equation}
        \label{eqn:VWP}
        \Psi(\bm{r},t) = Q(\omega t)\frac{u_{\bm{p}}}{\sqrt{2\e}}\exp\bigg[i\bigg(\p(t)\cdot \bm{r} - \int^tE(s)ds\bigg)\bigg]~,
\end{equation}
where $Q(\omega t)$ is the matrix
\begin{equation}
        \label{eqn:QMVA}
        Q(\omega t) = 1 - \frac{m\xi}{2\eta}(\gamma^0 -\gamma^3)(\,\gamma a(\omega t)\,)~,
\end{equation}
$\p(t) = \big(p_x,p_y,p_L(t)\big)$ is the quasimomentum and $E(s)$ is the quasienergy.  We call $p_L(t)$ the longitudinal quasimomentum. Further applying Eqs. (\ref{eqn:PZE}) to the Taylor expansion permits us to present the longitudinal quasimomentum and quasienergy as
\begin{align}
        p_{L\pm}(t) & = \frac{m^2 + \big( \bm{\kappa} \mp \xi m\bm{a}(\omega t) \,\big)^2-\eta^2}{2\eta}~,\label{eqn:LQM}
        \\
        E_{\pm}(t) & = \frac{m^2+ \big( \bm{\kappa} \mp \xi m\bm{a}(\omega t) \,\big)^2+\eta^2}{2\eta}~, \label{eqn:QE}
\end{align}
where $\bm{\kappa} = \big(p_x,p_y,0\big)$. Again for the sake of clarity we have suppressed the labels $\pm$ in all variables on the right-hand sides of these equations, except in front of $\xi$ where it is essential.

The mean quasienergy $\bar{E}$ can then be derived from Eq.(\ref{eqn:QE}), via its definition
\begin{equation}
        \label{eqn:MQE}
        \bar{E}_\pm = \frac{1}{T}\int_0^T \! E_\pm (s)\,ds.
\end{equation}
Note that from Eq. (\ref{eqn:VWP})  we then may write
$\Psi(\boldsymbol{r},t) = e^{-i\bar{E}t}\varphi(\boldsymbol{r},t)$ just as we did in Eq.(\ref{eqn:PSI}), where $\varphi(\boldsymbol{r},t)$ is a periodic function of $t$ with period $T$.

The advantage of the vicinal approximation is that the time and space variables are decoupled in the wave function (\ref{eqn:VWP}), greatly simplifying evaluation of the integrals in the amplitude (\ref{eqn:PAGF}). That is, the amplitude becomes
\begin{equation}
        \M = \frac{1}{T}\int_0^T \mathcal{K}(t)\,\mathcal{V}(t)\,\exp\big(i\SA(t)\big)\,dt,\label{eqn:PA}
\end{equation}
where the factor ${\mathcal K}(t)$ includes the spinor variables
\begin{equation}
        \label{eqn:SFD}
        {\mathcal K}(t) = \frac{1}{2\,(\e_-\e_+)^{1/2}}\,\big(\bar{u}_{\bm{p}_-}\,{\bar{Q}}_-(\omega t)\,\gamma^0\,{Q}^{*}_+\,(\omega t)\,\tilde{u}_{\bm{p}_+}^*\big)~,
\end{equation}
$\mathcal{V}(t)$ is the Fourier transform of the Coulomb potential
\begin{equation}
\label{eqn:FTCP}
        \mathcal{V}(t) = -\frac{4\pi Z \alpha}{|\p_+(t) + \p_-(t)|^2}~,
\end{equation}
and $\SA(t)$ is the contribution of the total quasienergy $E(t) = E_+(t) + E_-(t)$ to the action
\begin{equation}
        \label{eqn:AD}
        \SA(t) = \int^{t}E(s)ds~.
\end{equation}

\section{Linear Polarization}
\label{sec:LP}
Consider an electromagnetic plane wave with linear polarization
\begin{equation}
\label{eqn:LPD}
        \mathbf{a}(\Phi) = \big(0,-\sin{\Phi},0,0\big).
\end{equation}
In this section we write the lepton momenta as $p_{\pm} = \left(\e_{\pm},\kp{\E}{\pm},\kp{\B}{\pm},p_{z\pm}\right) = (\e_{\pm},\bm{p}_{\pm})$. The $\E$ and $\B$ superscripts denote the electric and magnetic field directions respectively, which correspond to the $x$ and $y$ coordinate directions. Further,  $\bm{\kappa}_{\pm}= \left(\kp{\E}{\pm},\kp{\B}{\pm}\right)$ is the lepton momentum transverse to the wavevector.

\subsection{Probability}
\label{sec:LPP}
For the polarization given by Eq. (\ref{eqn:LPD}) the total quasienergy found from Eq. (\ref{eqn:QE}) may be rewritten in the simple form
\begin{equation}
        \label{eqn:STQE}
        E(t) = \lambda\Big(\left(\sin{\omega t} + \beta\right)^2 + \chi^2\Big),
\end{equation}
in which
\begin{align}
        \lambda & = \frac{\xi^2m^2}{2}\left(\frac{1}{\eta_+} + \frac{1}{\eta_-}\right), \label{eqn:LD}\\
        \beta & =  \frac{1}{m\xi}\frac{\kp{\E}{+}\eta_{-} - \kp{\E}{-}\eta_{+}}{\eta_{-} + \eta_{+}},\label{eqn:BD}\\
        \chi & = \frac{1}{m\xi}\!\left(\! m^2 + \eta_+\eta_- + \frac{\bm{\kappa}_{+}^2\eta_{-} +\bm{\kappa}_{-}^2\eta_{+}}{\eta_{-} + \eta_{+}} - (m\xi\beta)^2\! \right)^{\!\!1/2}\!\!\!.\label{eqn:CD}
\end{align}
From Eq. (\ref{eqn:STQE}), the mean total quasienergy is clearly
\begin{equation}
        \label{eqn:MSTQE}
        \bar{E} = \lambda\big(1/2 + \beta^2 + \chi^2\big).
\end{equation}
The mean quasienergy has a minimum when
$\bm{\kappa}_+ =\bm{\kappa}_-=\bm{0}$, $\eta_+=\eta_-=m(1+\xi^2/2)^{1/2}$. At this minimum
\begin{equation}
        \label{eqn:LPEM}
        \bar{E} = 2m(1+\xi^2/2)^{1/2}\simeq \sqrt{2}m\xi~,
\end{equation}
in the tunneling regime. This minimum is the threshold energy required for pair creation, and it grows linearly with $\xi$:
$\bar{E}= 2 m(1+\xi^2/2)^{1/2} \simeq \sqrt{2} m\xi$.

Also from Eq.(\ref{eqn:STQE}) one finds the action (\ref{eqn:AD}) is
\begin{equation}
        \label{eqn:SA}
        \SA(t) = \frac{m}{\omega}\left(\frac{\bar{E}}{m}\omega t - \frac{\lambda}{2m}\cos{\omega t}\left(\sin{\omega t} + 4\beta\right)\right).
\end{equation}
Eq. (\ref{eqn:AR}) ensures that the coefficient $m/\omega$ in front of this expression is large. Consequently, the integral over time in Eq. (\ref{eqn:PA}) can be evaluated by the saddle point method. Labeling the saddle points as $t_j$, $j = 1,2,\ldots$, we find for the amplitude
\begin{equation}
        \label{eqn:PASPMLP}
        \M=\frac{1}{T}\sum_{j}{\mathcal K}(t_j)\mathcal{V}(t_j)\left(\frac{2\pi }{i E^{\prime}(t_j)}\right)^{1/2} e^{i\mathcal{S}(t_j)}~,
\end{equation}
where the prime denotes differentiation with respect to $t$.

The saddle points must satisfy the following condition:
\begin{equation}
        \label{eqn:SPD}
        \SA^{\prime}(t_j)=E(t_j) = 0~.        
\end{equation}
In other words, the pair is produced at the moment it has zero quasienergy. From Eq. (\ref{eqn:STQE}) this condition is equivalent to
\begin{equation}
        \label{eqn:TSPS}
        \sin{\omega t_j} = -\beta \pm i\chi.
\end{equation}
We are interested only in the saddle points with real part of $t_j$ lying in $[0,T]$, so there are only two pairs of complex solutions of Eq. (\ref{eqn:TSPS}): one solution of each pair lies in the upper-half complex plane and another in the lower-half complex plane. Since pair creation requires absorption of energy from the laser field, the theory of adiabatic transitions \cite{Dyk60,LL3-01} specifies that only the saddle points lying in the upper-half complex plane contribute to pair creation. Hence we have two saddle points
\begin{align}
\omega t_1 & = \sin^{-1}(\beta + i\chi) + \pi\mod 2\pi~,  \label{eqn:TSPLP1}\\
\omega t_2 & = \sin^{-1}(-\beta + i\chi)~~~\mod 2\pi~. \label{eqn:TSPLP2}
\end{align}
These saddle points have the properties that
\begin{align}
\Im(\omega t_1)  & = \Im(\omega t_2) > 0 \label{eqn:IPTSPLP}\\
\Re(\omega t_2) & = \pi - \Re(\omega t_1) \mod 2\pi.  \label{eqn:RPTSPLP}
\end{align}
Examination of Eq. (\ref{eqn:SA}) reveals that the action $\SA(t_j)$ consequently satisfies
\begin{align}
\Im\left[\SA(t_1)\right] & = \Im\left[\SA(t_2)\right] > 0\label{eqn:APLP1}\\
\Re\left[\SA(t_2)\right] & = \frac{\bar{E}\pi}{\omega} - \Re\left[\SA(t_1)\right] \mod 2\pi. \label{eqn:APLP2}
\end{align}

The pre-exponential terms of the sum in Eq. (\ref{eqn:PASPMLP}) prove be same for either saddle point. This fact combined with  Eqs. (\ref{eqn:APLP1}) and (\ref{eqn:APLP2}) permits us to express the amplitude in terms of only one saddle point, for example $t_1$. By substituting Eqs. (\ref{eqn:APLP1}) and (\ref{eqn:APLP2}) into Eq. (\ref{eqn:PASPMLP}), the probability becomes
\begin{align}
        |\M|^2
        &= \frac{\omega^2}{\pi}\frac{|{\mathcal K}(t_1)|^2\mathcal{V}(t_1)^2}{|E^{\prime}(t_1)|}\exp\big(-2\Im\left[\SA(t_1)\right]\big) \notag\\
        &  \times \bigg\{1 + \cos\bigg[\frac{1}{\omega}\bigg(\bar{E}\pi - 2\omega\Re\left[\SA(t_1)\right]\bigg)\bigg]\bigg\}. \label{eqn:PLP}
\end{align}
The phase within the cosine is large, being $\propto m/\omega$, so that as the lepton momenta vary slightly we expect the cosine term to oscillate very rapidly. As a result, we may neglect the contribution of this term to the integral in Eq. (\ref{eqn:TPCRGF}), and hence to any of the spectra or rates discussed in the present work. We therefore ignore this term henceforth.

\subsection{Pair Creation Rate}
\label{sec:LPPCR}
It now follows from Eq. (\ref{eqn:PLP}) that the pair creation rate (\ref{eqn:TPCRGF}) is
\begin{equation}
        \label{eqn:PCRIE}
        W = 2\omega\sum_{s_{\pm}}\int\frac{|\mathcal{K}(t_1)\mathcal{V}(t_1)|^2}{|E^{\prime}(t_1)|}
        e^{\left(-2\Im\left[\SA(t_{1})\right]\right)}\frac{d^3\bm{p}_-}{(2\pi)^3}\frac{d^3\bm{p}_+}{(2\pi)^3}.
\end{equation}

Let us find the minimum of $\Im [\SA(t_1)]$ as a function of electron and positron momenta.  For this purpose, it is convenient to define $P = (\kp{\E}{+},\kp{\B}{+},\eta_+,\kp{\E}{-},\kp{\B}{-},\eta_-)$ and to write all functions as explicit functions of both $P$ and $t$, noting that the saddle point $t_1 = t_1(P)$. The minimum of $\Im[\SA(t_1,P)]$ then satisfies $\nabla_P\Im\big[\SA(t_0,P_0)\big] = \bm{0}$ and corresponds to the most probable configuration of electron and positron momenta.

It is instructive to compare the current problem to the well-known static case for homogeneous electric and magnetic fields, where the fields are orthogonal and equal in magnitude: i.e. $\bm{E}\cdot\bm{B} = 0,~ \bm{B}^2 = \bm{E}^2$. In this case any charged particle would accelerate mostly in the $\bm{E}\times\bm{B}$ direction regardless of the sign of its charge. We may expect therefore that the minimum of $\Im [\SA(t_1,P)]$ is achieved when the electron and positron move along the $z$-direction.

By considering the partial derivatives of $\Im[\SA(t_0,P)]$ with respect to $\kp{\E}{\pm}$ and $\kp{\B}{\pm}$, it is straightforward to verify that at the minimum of $\Im[\SA(t_1,P)]$,
\begin{equation}
        \label{eqn:LPMA}
        \bm{\kappa}_+ = \bm{\kappa}_- = \bm{0}.
\end{equation}
Further, from the symmetry of the problem, we expect that the electron and positron momenta should be equal at the minimum of $\Im[\SA(t_1,P)]$, so we expect  $\eta_+ = \eta_-$, since $\eta$ defines the z-direction momentum $p_z$ by Eqs. (\ref{eqn:PZE}). This condition together with Eq. (\ref{eqn:LPMA}) permits the form of the quasienergy (\ref{eqn:QE}) to be simplified, and so   the action (\ref{eqn:AD}) becomes
\begin{equation}
        \label{eqn:AAM}
        \SA(t_1,\eta) =\frac{1}{\eta} \int^{t_1}\big\{m^2\big[1 +\xi^2\sin^2(\omega t')\big] + \eta^2\big\}~dt',
\end{equation}
at the minimum of $\Im[\SA(t_1,P)]$.

Applying Eq. (\ref{eqn:LPMA}) to Eqs. (\ref{eqn:BD}), (\ref{eqn:CD}), and (\ref{eqn:TSPS}), we also find that $\beta =0$, $\chi = (m^2 + \eta^2)^{1/2}/m\xi$ and $\sin(\omega t_1) = i\chi$ respectively. We presume that at the minimum $\eta \sim m$ (we verify this in Eq. (\ref{eqn:LPMSP}) below). Then for $\xi \gg 1$, we find
\begin{equation}
        \label{eqn:TSPAM}
        \omega |t_1| = \frac{1}{\xi}\bigg(1+ \frac{\eta^2}{m^2}\bigg)^{1/2} \ll 1.
\end{equation}
It follows from the inequality that $\sin(\omega t') \simeq \omega t'$ in Eq. (\ref{eqn:AAM}). After applying Eq. (\ref{eqn:TSPAM}) to the action (\ref{eqn:AAM}) we then have
\begin{equation}
        \label{eqn:LPAEM}
        \SA(\eta) = i\frac{2}{3\F}\frac{(m^2 + \eta^2)^{3/2}}{m^2\eta}.
\end{equation}
The imaginary part of this expression is minimal at $\eta = m/\sqrt{2}$, so $\Im[\SA(t_1,P)]$ is minimal at
\begin{equation}
        \label{eqn:LPMSP}
        P_0 = \bigg(0,0,\frac{m}{\sqrt{2}},0,0,\frac{m}{\sqrt{2}}\bigg)
\end{equation}
with minimum
\begin{equation}
        \Im[\SA(t_1,P_0)] = \frac{\sqrt{3}}{\F}.
\end{equation}
 
It is important to note that from Eq. (\ref{eqn:LQM}) the quasimomentum at the minimum is
\begin{equation}
        \label{eqn:LPQMSP}
                \p(t_1)_{\pm} = \bigg(0,0,-\frac{m}{\sqrt{2}}\bigg).
\end{equation}
Equation (\ref{eqn:LPQMSP}) shows that the typical momenta in the Coulomb matrix element of Eq. (\ref{eqn:PAGF}) are $|\p| \sim m$. This indicates that the distance from the nucleus within which the pair is created is comparable to the Compton radius. I.e. $r\lesssim 1/|\p|\sim 1/m$. This estimate supports the validity of the vicinal approximation introduced in Section \ref{sec:TFVA}.

We may expand $\Im[\SA(t_1,P)]$ in a Taylor series in powers of $P$ about the minimum $P_0$. This expansion may be calculated by using Eqs. (\ref{eqn:STQE})-(\ref{eqn:SA}) and (\ref{eqn:TSPS}). The result to second order in $P$ is
\begin{align}
        \Im[\SA(t_1,P)]
        & = \frac{\sqrt{3}}{\F} + \frac{1}{2}(P-P_0)^TH_{P_0}(P-P_0) \notag\\
        & \equiv \frac{\sqrt{3}}{\F}\big[1 + \mathcal{A}(P)\big]~,         \label{eqn:LPAEF}
\end{align}
where $H_{P_0}$ is the $6\times6$ Hessian matrix of second order partial derivatives of $\Im[\SA(t_0,P)]$ at $P = P_0$ and $(P-P_0)^T$ denotes the transpose of $(P-P_0)$. In the regime $\xi \gg 1$, this matrix is
\begin{equation}
        \label{eqn:LPHM}
        H_{P_0} =  \frac{\sqrt{3}}{2m^2\F}\begin{bmatrix}
        1 + \frac{1}{4\xi^2} & 0 & 0 & 1 - \frac{3}{4\xi^2}& 0 & 0\\
        0 & 2 & 0 & 0 & 0 & 0\\
        0 & 0 & \frac{7}{3} & 0 &0 & \frac{1}{3}\\
        1 - \frac{3}{4\xi^2} & 0 & 0 & 1 + \frac{1}{4\xi^2}& 0 & 0\\
        0 & 0 & 0 &0 & 2  &0\\
        0 & 0 & \frac{1}{3} & 0 &0 &\frac{7}{3}
\end{bmatrix}.
\end{equation}
The higher order $1/\xi^2$ terms are included in the matrix so $H_{P_0}$ is not singular. From Eqs. (\ref{eqn:LPAEF}) and (\ref{eqn:LPHM}) we may thus write for $\xi \gg 1$
\begin{align}
        \mathcal{A}(P)
        & = \frac{7(\delta\eta_-^2 + \delta\eta_+^2) + 2\delta\eta_-\delta\eta_+}{12m^2} + \frac{(\kp{\B}{-})^2 + (\kp{\B}{+})^2}{2m^2}\notag\\
        & \quad + \frac{(\kp{\E}{-} + \kp{\E}{+})^2}{4m^2} + \frac{(\kp{\E}{-} - \kp{\E}{+})^2}{8\xi^2m^2},        \label{eqn:LPAP}
\end{align}
where $\delta\eta_{\pm} = \eta_{\pm} - m/\sqrt{2}$ is the variation of $\eta_{\pm}$.

The factor $\sqrt{3}/\F$ in Eq. (\ref{eqn:LPAEF}) is large since $\F \ll 1$, so we
can apply the saddle point method in order to evaluate the integral in the pair creation rate (\ref{eqn:PCRIE}).
We now change integration variables in the integral (\ref{eqn:PCRIE}) from $d^3\bm{p}_-d^3\bm{p}_+$ to $d^6P$, so the minimum point $P_0$ is the only saddle point for the new integral. The differential pair creation rate thus becomes
\begin{equation}
        \label{eqn:DPCRGF}
        dW = \mathcal{B}\exp\bigg(-\frac{2\sqrt{3}}{\F}\mathcal{A}(P)\bigg)\frac{d^6P}{(2\pi)^6},
\end{equation}
where the coefficient
\begin{equation}
        \mathcal{B} = 2\omega\sum_{s_{\pm}}\frac{|\mathcal{K}(t_1)\mathcal{V}(t_1)|^2}{|E^{\prime}(t_1)|}
        \bigg(\frac{\partial p_{z}}{\partial\eta}\bigg)^2\bigg|_{P_0}\exp\bigg(-\frac{2\sqrt{3}}{\F}\bigg)~,
\end{equation}
and the saddle point $t_1$ is evaluated at $P_0$.

The only spin dependent term in $\mathcal{B}$ is $\mathcal{K}(t_1)$. In Appendix \ref{app:SF} we verify that the spin dependent factors give the following contribution to the creation rate
\begin{equation}
        \sum_{s_{\pm}}|\mathcal{K}(t_1)|^2 = \frac{8}{9}~.
\end{equation}
Further, Eqs. (\ref{eqn:LPQMSP}) and (\ref{eqn:FTCP}) permit us to calculate the matrix element of the Coloumb potential
\begin{equation}
        |\mathcal{V}(t_1)|^2 = \bigg(\frac{2\pi Z\alpha}{m^2}\bigg)^2~,
\end{equation}
and Eqs. (\ref{eqn:STQE}), (\ref{eqn:TSPAM}), and (\ref{eqn:LPMSP}) together produce $E^{\prime}(t_1) = i2\sqrt{3}\F m^2$ in the $\xi \gg 1$ regime. From Eq. (\ref{eqn:PZE}), the partial derivative $\partial p_z/\partial \eta = -3/2$ at the saddle point. Hence,
\begin{equation}
        \label{eqn:LPBC}
        \mathcal{B} = \frac{8\pi^2}{\sqrt{3}}\frac{Z^2\alpha^2\omega}{\F m^6}\,\exp\bigg(-\frac{2\sqrt{3}}{\F}\bigg)~.
\end{equation}
Straightforward integration of Eq. (\ref{eqn:DPCRGF}) produces the following result for the pair creation rate in the linear polarization case:
\begin{equation}
        \label{eqn:LPPCR}
        W =\frac{Z^2\alpha^2m}{\pi\sqrt{2}}\left(\frac{\F}{2\sqrt{3}}\right)^3
        \exp\bigg(-\frac{2\sqrt{3}}{\F}\bigg)~.
\end{equation}
We will check later (see Eq. \ref{eqn:VLPR}) that this formula agrees with the previously known results.

\subsection{Positron Spectrum}
\label{sec:LPPS}
Let us find the energy spectrum of the positrons. We may integrate the differential pair creation rate in Eq. (\ref{eqn:DPCRGF}) over the electron momenta $d\kp{\E}{-}d\kp{\B}{-}d\eta_-$, so that
\begin{equation}
        \label{eqn:DPCRIPM}
        dW =  W_+d\kp{\E}{+}d\kp{\B}{+}d\eta_+.
\end{equation}
Clearly $W_+$ represents the differential positron creation rate, as it is only a function of the positron momenta $\big(\kp{\E}{+},\kp{\B}{+},\eta_+\big)$. We then have
\begin{equation}
        \label{eqn:LPDPCR}
        W_+ = \mathcal{B}_{+}\,\exp\bigg(-\frac{\sqrt{3}}{\G}
        \mathcal{A}_{+}\big(\kp{\E}{+},\kp{\B}{+},\eta_+\big)\bigg)~,  
\end{equation}
where $\G \equiv \F\xi^2$ and the exponent and coefficient are respectively
\begin{gather}
        \mathcal{A}_{+}\big(\kp{\E}{+},\kp{\B}{+},\eta_+\big)  =\frac{(\kp{\E}{+})^2}{m^2} + \frac{(\kp{\B}{+})^2\xi^2}{m^2} +\frac{8(\delta\eta_+)^2\xi^2}{7m^2},\label{eqn:LPDPCRE}\\
        \mathcal{B}_{+}  = \frac{4}{(2\pi)^{5/2}\sqrt{7}}\frac{Z^2\alpha^2}{\xi m^2}\left(\frac{\F}{2\sqrt{3}}\right)^{3/2}\exp\left(-\frac{2\sqrt{3}}{\F}\right).
\end{gather}
The positron energy spectrum is $d W/d \bar{E}_+$, where $\bar{E}_+$ is the positron mean quasienergy
in the laser field. (An empirically important property of the created positrons is their energy outside the laser field.  If one assumes that the positron leaves the focus of the field adiabatically, then its energy outside the laser field is exactly its mean quasienergy $\bar{E}_+$.)

In order to find the positron energy spectrum, we must write $dW$ (\ref{eqn:DPCRIPM}) as a differential form which includes $d\bar{E}_+$, so that a change of variables is required. The transformation of $dW$ under a change of variables $\nu(y) = x$ from $x = (x_1,\ldots,x_n)$ to $y = (y_1,\ldots,y_n)$ for the saddle point $x_0 = \nu(y_0)$ reads
\begin{equation}
        \label{eqn:TSPM}
        dW = \mathcal{B_+}\big|J_{y_0}\nu\big|\exp\!
        \bigg(\!\!(y-y_0)^T\!\!(J_{y_0}\nu)^T\!\!H_{+}(J_{y_0}\nu)(y-y_0)\!\!\bigg)d^n\!y
\end{equation}
where $J_{y_0}\nu$ is the Jacobian matrix of $\nu$ evaluated at saddle point $y_0$, and $|J_{y_0}\nu|$ is its determinant. From Eq. (\ref{eqn:LPDPCRE}) the matrix $H_+$ is simply the diagonal $3\times 3$ matrix
\begin{equation}
H_+ = -\frac{\sqrt{3}}{\G m^2}\mbox{diag}\big\{1,\xi^2, 8\xi^2/7\big\}.
\end{equation}
From Eqs. (\ref{eqn:QE}), (\ref{eqn:MQE}), and (\ref{eqn:LPD}) and the fact that $\eta > 0$, one finds that
\begin{equation}
        \label{eqn:LPEMQE}
        \eta_+ = \bar{E}_+ - \big[\bar{E}_+^2 - \bm{\kappa}^2 - m^2(1 + \xi^2/2)\big]^{1/2}.
\end{equation}
Equation (\ref{eqn:LPEMQE}) defines the required coordinate transformation  $\nu_E(\kp{\E}{+},\kp{\B}{+}, \bar{E}_+) = (\kp{\E}{+},\kp{\B}{+},\eta+)$.  At the saddle point, $\bar{E}_{+0} = m(3+\xi^2)/2\sqrt{2} \simeq m\xi^2/2\sqrt{2}$, for $\xi \gg 1$. Substituting $\nu_E$ and the saddle point into Eq. (\ref{eqn:TSPM}) and then integrating over $\kp{\E}{+}$ and $\kp{\B}{+}$, one finds the positron energy spectrum
\begin{align}
        \frac{d W}{d\bar{E}_+}
        & = \frac{8}{(2\pi)^{3/2}\sqrt{7}}\frac{Z^2\alpha^2}{\xi^2}\left(\frac{\F}{2\sqrt{3}}\right)
        ^{5/2}\exp\left(-\frac{2\sqrt{3}}{\F}\right)\notag\\
        & \indent \times \exp\left[-\frac{32\sqrt{3}}{7\F\xi^4}\left(\frac{\bar{E}_+}{m} - \frac{\xi^2}{2\sqrt{2}} \right)^2\right]~.  \label{eqn:LPPES}
\end{align}
\subsection{Angular Distribution}
\label{sec:LPAD}

Let us find the positron angular distribution, $dW/d\Omega$, which indicates the direction of the mean velocity of the positron within the focus of the laser field. (If parameters of the field are known, one can extract from this information the angular distributions of the leptons outside the laser field.) We employ the spherical coordinates $(|\bm{p}_+|,\theta,\varphi)$ and $d\Omega = \sin\theta d\theta d\varphi$. In this coordinate system,
\begin{gather}
         \kp{\E}{+} = |\bm{p}_+|\cos\varphi\sin\theta~,\notag\\
        \kp{\B}{+} = |\bm{p}_+|\sin\varphi\sin\theta~, \label{eqn:LPSCT}\\
        \eta_+ = \left(m^2 + |\bm{p}_+|^2\right)^{1/2} - |\bm{p}_+|\cos\theta~,\notag
\end{gather}
which follows from Eq. (\ref{eqn:PZE}). These together define the transformation $\nu_{\Omega}(|\bm{p}_+|,\theta,\varphi) = (\kp{\E}{+},\kp{\B}{+},\eta_+)$. The saddle point in the spherical coordinates is located at $|\bm{p}_+| = m/2\sqrt{2}$ and $\theta = 0$. Note that the azimuthal angle $\varphi$ is not well-defined at this saddle point, so we keep all orders of $\varphi$ in the differential positron creation rate exponent $\mathcal{A}_+$.
After application of Eq. (\ref{eqn:TSPM}) and integration over $|\bm{p}_+|$ one finds
\begin{align}
        \frac{dW}{d\Omega}
        & = \frac{1}{16\pi^{2}\sqrt{2}}\frac{Z^2\alpha^2m}{\xi}\left(\frac{\F}{2\sqrt{3}}\right)
        ^2\exp\left(-\frac{2\sqrt{3}}{\F}\right)\notag\\
        & \times \exp\bigg(-\frac{\sqrt{3}}{8\G}\big(\cos^2\varphi + \xi^2\sin^2\varphi\big)\theta^2\bigg).\label{eqn:LPARS}
\end{align}
We assume in this derivation that the parameter $\G \equiv \F\xi^2$ is small, i.e. $\G \ll 1$, so that the angular distribution is integrable. The condition $\G \ll 1$ represents a special case of the tunneling regime, and it is equivalent to the additional inequality $m/\omega \gg \xi^3$ (see Eq. (\ref{eqn:TR})). The operational parameters of an XFEL may fall just inside this regime \cite{Rin01}.

\subsection{Photon Absorption Spectrum}
Let us find the spectrum of the number of absorbed photons in the pair creation process. In Section \ref{sec:TFCRPA} we used the adiabatic condition $\omega \ll m$ (\ref{eqn:AR}) to remove the sum over $n$ from the pair creation rate in Eq. (\ref{eqn:TPCR}). We now define the photon absorption spectrum as
\begin{equation}
        \label{eqn:NPCR}
         W_n = 2\pi\sum_{s_\pm}\int|\M|^2\delta
        (\bar{E} -  n\omega)\frac{d^3\bm{p}_+}{(2\pi)^3}\frac{d^3\bm{p}_-}{(2\pi)^3},
\end{equation}
where the total mean quasienergy $\bar{E} = \bar{E}_+ + \bar{E}_-$. Note that by comparison to the pair creation rate (\ref{eqn:TPCR}), $\sum_n W_n = W$, or $\int W_ndn =  W$ for $\omega \ll m$.

We now replace the Dirac delta function by its Fourier transform:
\begin{equation}
        \delta(\bar{E}-n\omega) = \frac{1}{2\pi}\int e^{i\tau \big(\bar{E}-n\omega\big)} d\tau.
\end{equation}
From Eq. (\ref{eqn:PLP}) we then write
\begin{align}
        W_n
        & = \frac{\omega^2}{\pi}\sum_{s_\pm} \! \int \! e^{-in\omega\tau}\frac{|\mathcal{K}(t_1)\mathcal{V}(t_1)|^2}{|E^{\prime}(t_1)|} \notag\\
        & \quad  \times \exp\Big\{i\tau\bar{E}-2\Im[\SA(t_{1})]\Big\}\frac{d^3\bm{p}_+}{(2\pi)^3}\frac{d^3\bm{p}_-}{(2\pi)^3}d\tau, \label{eqn:LPPASFT}
\end{align}

Just as shown in Section \ref{sec:LPPCR} for the integral in Eq. (\ref{eqn:PCRIE}), the integral over momenta in Eq. (\ref{eqn:LPPASFT}) may be evaluated by the saddle point method. From Eqs. (\ref{eqn:STQE})-(\ref{eqn:CD}), the $i\tau\bar{E}$ term in the exponent of Eq. (\ref{eqn:LPPASFT}) does not have a $1/\F$ factor, so that  when $\F \ll 1$ the saddle point $P_0$ is the same as it was in Eq. (\ref{eqn:LPMSP}). Upon application of the saddle point method, the $i\tau\bar{E}$ factor is instead expanded about $P_0$ to first order. Hence
\begin{equation}
        W_n = \frac{\omega}{2\pi}\mathcal{B}\int \! e^{i\tau[\bar{E}(P_0) - n\omega]} \exp\big[\mathcal{C}(P,\tau)\big]\frac{d^6P}{(2\pi)^6}d\tau,
\end{equation}
where the exponent
\begin{equation}
        \mathcal{C}(P,\tau) = i\tau\nabla_P\bar{E}(P_0)\cdot(P-P_0) - \frac{2\sqrt{3}}{\F}\mathcal{A}(P).
\end{equation}
With reference to the definition of $\mathcal{A}$ (\ref{eqn:LPAEF}) and employing some linear algebra, it can be shown that
\begin{align}
         W_n
        & = W\frac{\omega}{2\pi}\int \exp\bigg(i\tau(\bar{E}(P_0)-n\omega)\bigg)\notag\\
        & \times \exp\bigg(-\frac{\tau^2}{4}\big[\nabla_P\bar{E}(P_0)\big]^T
        H_{P_0}^{-1} \nabla_P\bar{E}(P_0)\bigg)d\tau.\label{eqn:GFPAS}
\end{align}
From Eqs. (\ref{eqn:LD})-(\ref{eqn:MSTQE}), (\ref{eqn:LPMSP}), and (\ref{eqn:LPHM}) we then have that for $\xi \gg 1$
\begin{align}
        W_n
        & = W\frac{\omega}{2\pi}\int \exp\bigg[i\tau\bigg(\frac{m\xi^2}{\sqrt{2}}-n\omega\bigg)\bigg]\notag\\
        & \times \exp\bigg(-\frac{\sqrt{3}}{128}\F m^2\xi^4\tau^2\bigg)d\tau.
\end{align}
Evaluating this integral produces the photon absorption spectrum in the tunneling regime:
\begin{align}
         W_n
        & = \frac{8}{(2\pi)^{3/2}\sqrt{3}}\frac{Z^2\alpha^2\omega}{\xi^2}\bigg(\frac{\F}{2\sqrt{3}}\bigg)
        ^{5/2}\exp\bigg(-\frac{2\sqrt{3}}{\F}\bigg)\notag\\
        & \quad  \times \exp\bigg[-\frac{32\omega^2}{\sqrt{3}\F m^2\xi^4}\bigg(n - \frac{m\xi^2}{\omega\sqrt{2}}\bigg)^2\bigg].\label{eqn:LPPAS}
\end{align}

\section{Circular polarization}
Consider an electromagnetic plane wave with circular polarization
\begin{equation}
        \label{eqn:CPP}
        \mathbf{a}(\Phi) = \big(0,-\sin\Phi,\cos\Phi,0\big).
\end{equation}
In this section it is natural to employ the cylindrical coordinate system, so we write the lepton momenta as $p = \big(\varepsilon,\kappa\cos\varphi,\kappa\sin\varphi,p^z\big) = \big(\epsilon,\bm{p}\big)$, and further $\bm{\kappa} = (\kappa\cos\varphi,\kappa\sin\varphi)$. Here the subscripts $\pm$ have been omitted for clarity.  It is also convenient to write the mean lepton angle and difference of lepton angles respectively as
\begin{equation}
        \label{eqn:CPA}
        \varphi \equiv \frac{\varphi_+ + \varphi _-}{2}\ \mbox{and}\ \phi  \equiv \frac{\varphi_+ - \varphi _-}{2}.
\end{equation}

\subsection{Probability}
\label{sec:CPP}
For the polarization given by Eq. (\ref{eqn:CPP}), the total quasienergy found from Eq. (\ref{eqn:QE}) may be written as
\begin{equation}
        \label{eqn:CPTQE}
        E(t) = \lambda\Big(\chi + \beta\sin(\omega t - \varphi -\sigma)\Big)~,
\end{equation}
where $\lambda$ is the same as in Eq. (\ref{eqn:LD}), and here
\begin{align}
        \beta & = \frac{2}{m\xi}\frac{|\eta_-\bm{\kappa}_+ - \eta_+\bm{\kappa}_-|}{\eta_+ + \eta_-}~,\label{eqn:BDCP}\\
  \chi & = \frac{1}{m^2\xi^2}\bigg(m^2(1 + \xi^2) + \eta_+\eta_- + \frac{\kappa_+^2\eta_- +
  \kappa_-^2\eta_+}{\eta_+ + \eta_-}\bigg)~,\label{eqn:CDCP}\\
  \sigma & = \tan^{-1}\bigg(\frac{\kappa_+\eta_- + \kappa_-\eta_+}{\kappa_+\eta_- - \kappa_-\eta_+}\tan\phi\bigg)~\label{eqn:SDCP}.
\end{align}

From Eq. (\ref{eqn:CPTQE}) the mean total quasienergy of the lepton pair is
\begin{equation}
        \label{eqn:MCPTQE}
        \bar{E} = \lambda\chi.
\end{equation}
The mean total quasienergy is minimal at  $\kappa_+ = \kappa_- = 0$ and $\eta_+ = \eta_- = m(1+\xi^2)^{1/2}$, with minimum value
\begin{equation}
                                \label{eqn:CPEM}
        \bar{E} = 2m(1+\xi^2)^{1/2} \simeq 2m\xi~,
\end{equation}
in the tunneling regime. This minimum is the threshold energy required for pair creation, and just as in the linear polarisation case, it grows linearly with $\xi$.

It also follows from Eq. (\ref{eqn:CPTQE}) that the action (\ref{eqn:AD}) is
\begin{equation}
        \label{eqn:CPSA}
        \SA(t) = \frac{m}{\omega}\bigg(\frac{\bar{E}}{m}\omega t + \frac{\lambda\beta}{m}\cos(\omega t - \varphi -\sigma)\bigg)~.
\end{equation}
Just as in Section \ref{sec:LP} the action has a coefficient $m/\omega$, which is large by Eq. (\ref{eqn:AR}). Hence the integral over time in Eq. (\ref{eqn:PA}) may be evaluated by the saddle point method.  For saddle points $t_j$ the amplitude $\M$ has the same form as in Eq. (\ref{eqn:PASPMLP}).

The saddle points must satisfy $E(t_j) = \SA'(t_j) = 0$, which here is equivalent to
\begin{equation}
        \label{eqn:TSPS2}
        \sin(\omega t_j - \varphi -\sigma) = -\frac{\chi}{\beta}.
\end{equation}
The right side of Eq. (\ref{eqn:TSPS2}) is real and we presume $|\chi/\beta| > 1$ (verified in Eq. \ref{eqn:TSPS3} below), so there is only a single pair of saddle points with real part in $[0,T]$: one lies in the upper half complex plane and the other in the lower half complex plane.  Just as in the linear polarization case we choose the saddle point lying in the upper half complex plane \cite{Dyk60,LL3-01}, and label it $t_0$. With reference to Eq. (\ref{eqn:PASPMLP}), since there is only a single saddle point the probability becomes
\begin{equation}
        \label{eqn:PCP}
        |\M|^2 =\frac{\omega^2}{2\pi}\frac{|\mathcal{K}(t_0)|^2\mathcal{V}(t_0)^2}{|E^{\prime}(t_0)|}\exp\big(-2\Im\left[\SA(t_0)\right]\big)~.
\end{equation}

\subsection{Pair Creation Rate}
\label{sec:CPPCR}
From Eq. (\ref{eqn:PCP}) the pair creation rate (\ref{eqn:TPCRGF}) is
\begin{equation}
        \label{eqn:PCRSSP}
         W = \omega\sum_{s_{\pm}}\int\frac{|\mathcal{K}(t_0)|^2\mathcal{V}(t_0)^2}{|iE^{\prime}(t_0)|}
        \,\,e^{-2\Im\left[\SA(t_{0})\right]}\,\,\frac{d^3\bm{p}_-}{(2\pi)^3}\frac{d^3\bm{p}_+}{(2\pi)^3}.
\end{equation}
This formula differs by a factor of $2$ from the rate in the linear polarization case (\ref{eqn:PCRIE}) since there is only a single saddle point $t_0$ here.

Let us find the minimum of $\Im[\SA(t_0)]$ as a function of the electron and positron momenta. For this purpose, it is convenient to define $P = \big(\kappa_-,\eta_-,\kappa_+,\eta_+,\phi,\varphi\big)$ and to write all functions as explicit functions of both $P$ and $t$, noting that the time saddle point $t_0 = t_0(P)$. The minimum of $\Im[\SA(t_0,P)]$ satifies $\nabla_P\Im[\SA(t_0,P)] = \bm{0}.$

The circularly polarized EM plane wave imparts angular momentum to the lepton pair, so we expect that $\kappa_+$ and $\kappa_-$ are not zero at the minimum.  By symmetry, we expect that at the minimum the electron and positron momenta are the same in size and the total momentum of the pair is conserved.  Hence we deduce that at the minimum:
\begin{eqnarray}
        \label{eqn:LMCP}
        \bm{\kappa}_+ &=& -\bm{\kappa}_-,
        \\ \label{eqn:LMCP2}
        \  \eta_+ &=& \eta_- = \eta.
\end{eqnarray}
Eq. (\ref{eqn:LMCP}) implies $\kappa_+ = \kappa_- = \kappa$. Moreover, if we choose the azimuthal angles so that $\varphi_- \in [0,2\pi]$ and $\varphi_+ \in [\pi,3\pi]$, then this condition means that
\begin{equation}
        \label{eqn:CPASPC}
        \varphi_ + = \varphi_- + \pi~.
\end{equation}
It follows by Eqs. (\ref{eqn:CPA}) that at a minimum $P_0$ of $\Im[\SA(t_0,P)]$
\begin{gather}
        \phi = \frac{\pi}{2}~,\ \mbox{and}\\
        \varphi \in \bigg[\frac{\pi}{2},\frac{5\pi}{2}\bigg].
\end{gather}        
Applying conditions (\ref{eqn:LMCP}) to Eqs. (\ref{eqn:BDCP}), (\ref{eqn:CDCP}), and (\ref{eqn:SDCP}) at this minimum, we also find that $\beta = 2\kappa/m\xi$, $\chi = [m^2(1+\xi^2) + \eta^2 + \kappa^2]/m^2\xi^2$ and $\sigma = \pm\pi/2$. The $\pm$ sign here is due to the fact that the angle $\sigma$ is not well-defined under conditions (\ref{eqn:LMCP}). The total quasienergy (\ref{eqn:CPTQE}) is then simplified, so that the action (\ref{eqn:AD}) becomes
\begin{equation}
        \label{eqn:CPMSPD}
        \SA(t_0) = \frac{1}{\eta}\int^{t_0}\!\!\big[m^2(1+ \xi^2) + \eta^2 + \kappa^2 \pm 2m\kappa\xi
        \cos(\omega t' - \varphi)\big]dt'
\end{equation}
at the minimum of $\Im[\SA(t_0,P)]$.

Consider now the partial derivatives with respect to $\eta$ and $\kappa$ of $\Im[\SA(t_0)]$ (\ref{eqn:CPMSPD}), which must be zero at the minimum. This produces two further equations, which may be divided by one another to derive the relation
\begin{equation}
        \label{eqn:CPDPD}
        \kappa^2 = m^2(1+\xi^2) - \eta^2~.
\end{equation}
Just as in the linear polarization case (Section \ref{sec:LPPCR}) we presume $\eta \sim m$. In fact, since the circularly polarized wave is equivalent to two distinct linearly polarized waves with particular phase and polarisation, and since by its definition (\ref{eqn:ED}) $\omega\eta$ is a frame invariant quantity, we expect $\eta = m/\sqrt{2}$ at the minimum. (This is verified below in Eq. \ref{eqn:CPMSP}.) Then from Eq. (\ref{eqn:CPDPD}), $\kappa = m(1/2 + \xi^2)^{1/2} \simeq m\xi$, for $\xi \gg 1$.

At the minimum Eq. (\ref{eqn:TSPS2}) now becomes
\begin{equation}
        \label{eqn:TSPS3}
        \cos(\omega t_0 -\varphi) = \mp\frac{\chi}{\beta} = \mp\frac{m^2(1 + 2\xi^2) + \eta^2}{2m^2\xi^2}~,
\end{equation}
from which it can be seen clearly that $|\chi/\beta| > 1$. For $\xi \gg 1$ we find
\begin{equation}
        \label{eqn:CPTSPI}
        |\omega t_0 - \phi| = \frac{1}{\xi}\bigg(1 + \frac{\eta^2}{m^2}\bigg)^{1/2} \ll 1~.
\end{equation}
It follows from the inequality in Eq. (\ref{eqn:CPTSPI}) that
\begin{equation}
        \cos(\omega t' - \phi) \simeq \mp\big[1 -(\omega t' -\phi)^2/2\big]~,
\end{equation}                
in the integral in Eq. (\ref{eqn:CPMSPD}). Applying Eq. (\ref{eqn:CPTSPI}) to the action in Eq. (\ref{eqn:CPMSPD}) we find that
\begin{equation}
        \SA(\eta) = i\frac{2}{3\F}\frac{(m^2 + \eta^2)^{3/2}}{m^2\eta} + \frac{\phi}{\omega\eta}\big[m^2(1 + 2\xi^2) + \eta^2\big].
\end{equation}        
The imaginary part of this expression (Eq.(\ref{eqn:LPAEM})) is minimal at $\eta = m/\sqrt{2}$. Importantly, this minimum is independent of the mean angle $\varphi$, so the minimum $P_0$ is degenerate in $\varphi$. This can be verified by noting the partial derivative of the action (\ref{eqn:CPSA}) with respect to $\varphi$ is purely real. Hence, $\Im[\SA(t_0,P)]$ is minimal at
\begin{equation}
        \label{eqn:CPMSP}
        P_0 = \bigg(m\xi,\frac{m}{\sqrt{2}},m\xi,\frac{m}{\sqrt{2}},\frac{\pi}{2},
        \left[\frac{\pi}{2},\frac{5\pi}{2}\right]\bigg)~,
\end{equation}        
with minimum value
\begin{equation}
        \Im[\SA(t_0,P_0)] = \frac{\sqrt{3}}{\F}~.
\end{equation}

The quasimomentum (\ref{eqn:LQM}) at the minimum is
\begin{equation}
        \label{eqn:QMSP}
        \p(t_0) = \bigg(m\xi\cos\varphi,m\xi\sin\varphi,-\frac{m}{2\sqrt{2}}\bigg)~,
\end{equation}
so in the circular polarisation the typical momenta in the Coulomb matrix element of Eq. (\ref{eqn:PAGF}) is $|\p| \sim m\xi$. This implies that for $\xi \gg 1$, the distance from the nucleus within which the lepton pair is created is much less than the Compton radius. I.e. $r \lesssim 1/|\p| \ll 1/m$. This estimate strongly supports the validity of the vicinal approximation introduced in Section \ref{sec:TFVA}.

From Eqs. (\ref{eqn:CPSA}) and (\ref{eqn:TSPS2}),  $\Im[\SA(t_0,P)]$ is independent of $\varphi$, so it is convenient to define $P = (R,\varphi)$ and write the imaginary part of the action as $\Im[\SA(t_0,R)]$. We now seek to expand $\Im[\SA(t_0,R)]$ in a Taylor series about $R_0$. This expansion is calculated using Eqs. (\ref{eqn:CPTQE})-(\ref{eqn:CPSA}) and (\ref{eqn:TSPS2}) and noting $\Im[t_0]$ is a function of $R_0$. To second order in $R$ abut $R_0$, the result of this expansion is
\begin{align}
        \Im[\SA(t_0,R)]
        & = \frac{\sqrt{3}}{\F} + \frac{1}{2}(R-R_0)^TH_{R_0}(R-R_0)\notag\\
        & \equiv \frac{\sqrt{3}}{\F}\big[1 + \mathcal{A}(R)\big], \label{eqn:CPSOE}
\end{align}
where $H_{R_0}$ is the $5\times 5$ Hessian matrix of second order partial derivatives of $\Im[\SA(t_0,R)]$ evaluated at $R_0$. In the regime $\xi \gg 1$,
\begin{equation}
        \label{eqn:CPHM}
        H_{R_0} = -\frac{\sqrt{3}}{m^2\F}\begin{bmatrix}
                1 & 0 &0 & 0  & 0\\
                0 & \frac{7}{6} &  0 & \frac{1}{6} & 0\\
                0 & 0& 1 & 0 & 0\\
                0 & \frac{1}{6} & 0 & \frac{7}{6} & 0\\
                0 & 0 & 0 & 0 & 2m^2\xi^2
                \end{bmatrix}.
\end{equation}
From Eqs. (\ref{eqn:CPSOE}) and (\ref{eqn:CPHM}), we then have
\begin{equation}
        \mathcal{A}(R) = \frac{7(\delta\eta_-^2 + \delta\eta_+^2) + 2\delta\eta_-\delta\eta_+}{12m^2} + \frac{\delta\kappa_+^2 +\delta\kappa_-^2}{2m^2} + \xi^2\delta\phi^2\!,
\end{equation}
where $\delta\eta = \eta - m/\sqrt{2}$, $\delta\kappa = \kappa - m\xi$,  and $\delta\phi = \phi - \pi/2$.

The factor $\sqrt{3}/\F$ in Eq. (\ref{eqn:CPSOE}) is large, so we may immediately apply the saddle point method to evaluate the integral in the pair creation rate (\ref{eqn:PCRSSP}). Clearly the saddle point is the minimum $P_0$. The degeneracy of the saddle point $P_0$ in its $\varphi$ component means that its contributions includes an integral over $\varphi$ from $\pi$ to $5\pi/2$. However, all terms in the integrand of Eq. (\ref{eqn:PCRSSP}) prove to be independent of $\varphi$, as expected by the cylindrical symmetry of the problem, so this integral merely contrbutes a factor of $2\pi$. Thus we have the differential pair creation rate
\begin{equation}
        \label{eqn:CPDPCR}
        dW = \mathcal{B}\exp\bigg(-\frac{\sqrt{3}}{\F}\mathcal{A}(R)\bigg)\frac{d^5R}{(2\pi)^5}~,
\end{equation}
where the coefficient
\begin{equation}
        \mathcal{B} = \omega\sum_{s_{\pm}}\frac{|\mathcal{K}(t_0)\mathcal{V}(t_0)|^2}{|E^{\prime}(t_0)|}
        \bigg(\frac{\partial p_{z}}{\partial\eta}\bigg)^2\bigg|_{R_0}\bigg|\frac{\partial(\varphi_+,\varphi_-)}{\partial(\phi,\varphi)}\bigg|e^{-\frac{2\sqrt{3}}{\F}}~,
\end{equation}        
and the saddle point $t_0$ is evaluated at $R_0$.
        
The only spin dependent term in $\mathcal{B}$ is $\mathcal{K}(t_0)$. In Appendix \ref{app:SF} we verify that
\begin{equation}
        \sum_{s_{\pm}}|\mathcal{K}(t_0)|^2 = \frac{2}{\xi^4}~.
\end{equation}
From Eqs. (\ref{eqn:FTCP}), (\ref{eqn:CPASPC}), and (\ref{eqn:QMSP}) one finds the Coulomb potential is
\begin{equation}
        |\mathcal{V}(t_0)|^2 = \bigg(\frac{2\pi Z\alpha}{m^2}\bigg)^2~.
\end{equation}
Equations (\ref{eqn:CPTQE}) and (\ref{eqn:CPTSPI}) in the regime $\xi \gg 1$ produce $E'(t_0) = i2\sqrt{3}\F m^2$, and by Eq. (\ref{eqn:PZE}), $\partial p_z/\partial \eta = -\xi^2$ at $R_0$. Further, the determinant $|\partial(\varphi_+,\varphi_-)/\partial(\phi,\varphi)| = 2$. We then have, see Eq.(\ref{eqn:LPBC}),
\begin{equation}
        \mathcal{B} = \frac{8\pi^2}{\sqrt{3}}\frac{Z^2\alpha^2\omega}{\F m^6}\,\exp\bigg(-\frac{2\sqrt{3}}{\F}\bigg)~.
\end{equation}
Straightforward integration of Eq. (\ref{eqn:CPDPCR}) produces the following result for the pair creation rate in the circular polarization case within the tunneling regime:
\begin{equation}
        \label{eqn:CPPCR}
         W = \frac{Z^2\alpha^2}{2\sqrt{\pi}}m\left(\frac{\F}{2\sqrt{3}}\right)^{5/2}\exp\bigg(-\frac{2\sqrt{3}}{\F}\bigg).
\end{equation}
This result is precisely the pair creation rate for a static crossed field combined with a Coulomb field \cite{Rit72,NN72}, which is expected as the electric field strength is constant in a circularly polarized plane wave. The result is also in agreement with the pair creation rate for a static homogeneous field combined with a Coulomb field \cite{BKS72}.
The pair creation rate for a linearly polarized EM plane wave may be derived from this result by assuming that the electric field in Eq. (\ref{eqn:CPPCR}) can be considered as time-dependent, $\E = \E\cos(\omega t)$. The linear pair creation rate will then be the average over one period of the plane wave. We thus have linear pair creation rate
\begin{align}
         W
        & = \frac{1}{T}\int_0^T\frac{Z^2\alpha^2}{2\sqrt{\pi}}m\left(\frac{\F}{2\sqrt{3}}
        |\cos(\omega t)|\right)^{5/2}\notag\\
        & \quad \quad \times \exp\bigg(-\frac{2\sqrt{3}}{\F|\cos(\omega t)|}\bigg)dt.
\end{align}
This can be evaluated by the saddle point method, since $\F \ll 1$, with saddle points at $t = 0,T$. Hence
\begin{align}
         W
        & = \frac{2}{T}\frac{Z^2\alpha^2m}{2\sqrt{\pi}}\left(\frac{\F}{2\sqrt{3}}\right)^{5/2}\int_{-\infty}^{\infty}\exp\bigg(-\frac{2\sqrt{3}}{\F}\frac{\omega^2t^2}{2}\bigg)dt\notag\\
        & = \frac{Z^2\alpha^2m}{\pi\sqrt{2}}\left(\frac{\F}{2\sqrt{3}}\right)^3\exp\left(-\frac{2\sqrt{3}}{\F}\right),\label{eqn:VLPR}
\end{align}
which is precisely the result derived above (\ref{eqn:LPPCR}). The linear polarization pair creation rate (\ref{eqn:LPPCR}) is therefore also in agreement with the static crossed field or homogeneous field results.

The form of the circular polarization pair creation rate (\ref{eqn:CPPCR}) agrees with the form of the result derived without use of the vicinal approximation \cite{AAMS03}, and both rates are also in exact agreement with the result derived by use of the polarization operator \cite{MMHJK06}.

\subsection{Positron Spectrum and Angular Distribution}
\label{sec:CPPS}
Let us find the spectra of the positrons. We employ the same approach to find these spectra as done for the linear polarization case in Section \ref{sec:LPPS}. In order to find the differential positron creation rate $W_+$ we integrate the differential pair creation rate in Eq. (\ref{eqn:CPDPCR}) over $d\kappa_-d\eta_-d\phi$, and write
\begin{equation}
        dW = W_+d\kappa_+d\eta_+.
\end{equation}
Note that by definition (\ref{eqn:CPA}) the angle $\phi$ describes the relative orientation of the electron and positron momenta, so we must integrate over this variable. We then have
\begin{equation}
        W_+ = \mathcal{B_+}\exp\bigg(-\frac{\sqrt{3}}{\F}\mathcal{A}_+(\kappa_+,\eta_+)\bigg)~,
\end{equation}
where the exponent and coefficient are respectively
\begin{gather}
        \mathcal{A}_+ = \frac{\delta\kappa^2}{m^2} + \frac{8\delta\eta^2}{7m^2}~,\label{eqn:CPEDPCR}\\
        \mathcal{B}_+ = \frac{2}{(2\pi)^{5/2}\sqrt{7}}\frac{Z^2\alpha^2}{m}\bigg(\frac{\F}{2\sqrt{3}}\bigg)\exp\bigg(-\frac{2\sqrt{3}}{\F}\bigg).
\end{gather}

We may write down the transformation of $dW$ under a general change of variable, as was done in Eq. (\ref{eqn:TSPM}). Such a change of variable is required to find the positron energy spectrum or the positron angular distribution. From Eq. (\ref{eqn:CPEDPCR}) the matrix $H_+$ for the circular polarization case is the diagonal $2\times 2$ matrix
\begin{equation}
        H_+ = -\frac{\sqrt{3}}{\F m^2}\mbox{diag}\big\{1,8/7\big\}.
\end{equation}

The positron energy spectrum is $dW/d\bar{E}_+$. From Eqs. (\ref{eqn:QE}), (\ref{eqn:MQE}), and (\ref{eqn:LPD}) and the fact that $\eta > 0$, we have
\begin{equation}
        \eta_+ = \bar{E}_+ - \big[\bar{E}_+^2 -\kappa_+^2 -m^2(1+\xi^2)\big]^{1/2}.
\end{equation}
This defines the transformation $\nu_E(\kappa_+,\bar{E}_+) = (\kappa_+,\eta_+)$, which permits $dW$ to be expressed in terms of the differential $d\bar{E}_+$. At the saddle point $P_1$, $\bar{E}_{+0} = m(3 + 4\xi^2)/2\sqrt{2} \simeq \sqrt{2}m\xi^2$, for $\xi \gg 1$. Substituting the coordinate transformation $\nu_E$ and the saddle point into Eq. (\ref{eqn:TSPM}) and then integrating over $\kappa_+$, one finds the positron energy spectrum
\begin{align}
        \frac{d  W}{d \bar{E}_+}
        & = \frac{1}{2\pi\sqrt{7}}\frac{Z^2\alpha^2}{\xi^2}\bigg(\frac{\F}{2\sqrt{3}}\bigg)^2\exp\bigg(-\frac{2\sqrt{3}}{\F}\bigg)\notag\\
        & \quad \times \exp\bigg[-\frac{2\sqrt{3}}{7\F\xi^4}\bigg(\frac{\bar{E}_+}{m} - \xi^2\sqrt{2}\bigg)^2\bigg]~.\label{eqn:CPPES}
\end{align}

Due to the cylindrical symmetry in the circular polarization case, the angular distribution here is simply $dW/d\theta$, where $\theta$ is the polar angle. In the spherical coordinates $(|\bm{p}_+|,\theta)$
\begin{gather}
        \kappa_+ = |\bm{p}_+|\sin\theta~,\notag\\
        \eta_+ = \big(m^2 + |\bm{p}_+|^2\big)^{1/2} - |\bm{p}_+|\cos\theta
\end{gather}
which follows from Eq. (\ref{eqn:PZE}). These relations define the coordinate transform $\nu_{\theta}(|\bm{p}_+|,\theta) = (\kappa_+,\eta_+)$, and the saddle point in spherical coordinates is located at $|\bm{p}_+| \simeq m\xi^2/2$, $\theta = \sqrt{2}/\xi$ for $\xi \gg 1$. After application of Eq. (\ref{eqn:TSPM}) and integration over $|\bm{p}_+|$, we have
\begin{align}
        \frac{d W}{d\theta}
        & = \frac{Z^2\alpha^2m\xi}{(2\pi)\sqrt{7}}\bigg(\frac{\F}{2\sqrt{3}}\bigg)^2\exp\bigg(-\frac{2\sqrt{3}}{\F}\bigg)\notag\\
        & \quad \times \exp\bigg[-\frac{2\sqrt{3}\xi^2}{7\F}\bigg(\theta - \frac{\sqrt{2}}{\xi}\bigg)^2\bigg].\label{eqn:CPARS}
\end{align}  

\subsection{Photon Absorption Spectrum}
Just as in the linear polarization case, the photon absorption spectrum in the circular polarization case is (see Eq.  (\ref{eqn:GFPAS}))
\begin{align}
         W_n
        & = W\frac{\omega}{2\pi}\int \exp\bigg(i\tau(\bar{E}(R_0)-n\omega)\bigg)\notag\\
        & \times \exp\bigg(\frac{-\tau^2}{4}\big[\nabla_R\bar{E}(R_0)\big]^T\ H_{R_0}^{-1}\ \nabla_R\bar{E}(R_0)\bigg)d\tau.
\end{align}
From Eqs. (\ref{eqn:LD}), (\ref{eqn:BDCP})-(\ref{eqn:MCPTQE}), (\ref{eqn:CPMSP}), and (\ref{eqn:CPHM}) one then finds that for $\xi \gg 1$
\begin{align}
        W_n
        & = W\frac{\omega}{2\pi}\int \exp\bigg(i\tau(2\sqrt{2}m\xi^2 - n\omega)\bigg)\notag\\
        & \times \exp\bigg(-\frac{\sqrt{3}}{8}\F m^2\xi^4\tau^2\bigg)d\tau.
\end{align}
Evaluating this integral produces the photon absorption spectrum:
\begin{align}
         W_n
        & = \frac{1}{2\pi\sqrt{3}}\frac{Z^2\alpha^2\omega}{\xi^2}\bigg(\frac{\F}{2\sqrt{3}}\bigg)
        ^2\exp\bigg(-\frac{2\sqrt{3}}{\F}\bigg)\notag\\
        & \quad \times \exp\bigg[-\frac{2\omega^2}{\sqrt{3}\F m^2\xi^4}
        \bigg(n-2\sqrt{2}\frac{m\xi^2}{\omega}\bigg)^2\bigg].\label{eqn:CPPAS}
\end{align}

\section{Elliptic polarization}
Consider an EM wave with the elliptic polarization
\begin{eqnarray}
\label{Aell}
\mathbf{a}=\big(0,-\sin \Phi,\,b\,\cos\Phi \big)~.
\end{eqnarray}
The parameter $b$, which is presumed to satisfy $0 \le  b \le1$,
measures the ellipticity: $b=0$ gives the linear polarization Eq.(\ref{eqn:LPD}), $b=1$ corresponds to the circular polarization Eq.(\ref{eqn:CPP}).  We verified that the probability of the pair creation achieves its maximum when
the electron and positron momenta satisfy
\begin{eqnarray}
\label{kap}
{\boldsymbol \kappa}_{-}&=&-{\boldsymbol \kappa}_{+}~, 
\\ \label{eta}
\eta_-&=&\eta_+=\eta~.
\end{eqnarray}
We already know that this condition holds for the linear and circular polarizations, as is discussed in and after Eq.(\ref{eqn:LPMA}), and Eq.(\ref{eqn:LMCP}). Further analyses reveals that
\begin{eqnarray}
\label{ka}
\kappa&=&b\, \left(\xi^2+1/2\right)^{1/2}\!m\simeq b \, \xi \,m~,
\\ \label{et}
\eta&=&\frac{m}{\sqrt{2}}~.
\end{eqnarray}
Here $\kappa=|{\boldsymbol \kappa}_-|=|{\boldsymbol \kappa}_+|$.
Calculating the behavior of the imaginary part of the action Eq.(\ref{eqn:AD}) considered a function of all momenta in the vicinity of its minimum we find
\begin{eqnarray}
\label{ImS}
\Im [\SA ] &=& \frac{\sqrt{3}}{\F}\big[\,1 + \mathcal{A}(P)\,\big]~,
\\ \nonumber
\mathcal{A}(P)&=& 
         \frac{7(\delta\eta_-^2 + \delta\eta_+^2) + 2\delta\eta_-\delta\eta_+}{12m^2} +
          \frac{(\delta \kp{\B}{-})^2 + (\delta \kp{\B}{+})^2}{2m^2} 
          \\  \label{A}
           &+& \frac{(\delta \kp{\E}{-} + \delta \kp{\E}{+})^2}{4m^2} + 
           \frac{1-b^2}{ 8\xi^2 } 
           \frac{( \delta\kp{\E}{-} - \delta \kp{\E}{+})^2 }{ m^2 }~.       
\end{eqnarray}
Here deviations of momenta from the values, which satisfy Eqs.(\ref{ka}),(\ref{et}) is implied.

Calculations, which are similar to the ones discussed previously for the linear and circular cases give the total rate $W_\mathrm{ell}$ of the pair production for a general elliptic polarization. It can be conveniently presented using a coeffecient $k_\mathrm{ell}$, which distinguishes it from a rate for the  linear polarization $W_\mathrm{lin}$, which is given in Eq.(\ref{eqn:LPPCR})
\begin{eqnarray}
\label{Well}
W_\mathrm{ell}&=&k_\mathrm{ell}\,W_\mathrm{lin}~,
\\ \label{kell}
k_\mathrm{ell}&=&\frac{1}{ (1-b^2)^{1/2} } \,\mathrm{erf}\left[
\frac{\pi}{2} \Bigg( \frac{\sqrt{3}(1-b^2)}{\F}\Bigg)^{\!1/2}\right]\,.\quad
\end{eqnarray}
Here $\mathrm{erf}(x)$ is a conventional error-function
\begin{equation}
\label{erf}
\mathrm{erf}(x)=\frac{2}{\sqrt{\pi}}\int_0^x \exp(-z^2)\,dz~.
\end{equation}
For $b=0$ one finds that $k_\mathrm{ell}\simeq 1$ (since $\F\ll 1$), which means that Eq.(\ref{Well}) correctly reproduces the rate for the linear polarization. For $b\rightarrow 1$, Eq.(\ref{kell}) gives $k_\mathrm{ell}=(  \pi \sqrt{3} / \F  )^{1/2}$. Substituting this result in Eqs.(\ref{Well}), (\ref{eqn:LPPCR}) one reproduces the rate for the circular polarization Eq.(\ref{eqn:CPPCR}). 
The elliptic polarization was discussed previously by Milstein {\it et al.} \cite{MMHJK06} presuming that the polarization is not close to the circular polarization, i.e. $b$ deviates significantly from 1. In that case, the error-function is close to unity and $k_\mathrm{ell}\simeq 1/\sqrt{1-b^2}$, which agrees with Ref. \cite{MMHJK06}.
An advantage of Eq.(\ref{Well}) is that it describes any polarization, without any restrictions.

Eqs.(\ref{ImS}), (\ref{A}) allow one to consider the spectral and angular distributions of the lepton pair for the elliptic polarization, though we will not dwell on this issue in this work.

\section{Discussion}
\subsection{Above-Threshold Pair Creation}

The positron energy and photon absorption spectra are in essence non-normalised probability distribution functions, which describe the probability of pair creation as a function of positron energy and number of absorbed photons respectively.   The asymptotic expressions obtained for the positron energy spectra (Eqs. (\ref{eqn:LPPES}) and (\ref{eqn:CPPES})) and photon absorption spectra (Eqs. (\ref{eqn:LPPAS}) and (\ref{eqn:CPPAS})) are all in Gaussian form. Since the maximum of a Gaussian probability distribution function is its expectation value, we may immediately write down the expected positron energies and photon absorption numbers, which are
\begin{align}
        \langle\bar{E}_+\rangle_{lin} & = \frac{m\xi^2}{2\sqrt{2}}, & \langle\bar{E}_+\rangle_{circ} & = \quad \sqrt{2}m\xi^2, \label{eqn:PEM}\\
        \langle n \rangle_{lin} & = \frac{m\xi^2}{\omega\sqrt{2}} & \langle n \rangle_{circ} & =  \frac{2\sqrt{2}m\xi^2}{\omega},
\end{align}
where the subscripts `lin' and `circ' denote the linear and circular polarization respectively.

The threshold quasienergy $\bar{E}_{0}$ required for pair creation was derived in Eqs. (\ref{eqn:LPEM}) and (\ref{eqn:CPEM}) for each polarization case. By symmetry, $\bar{E}_0 = 2\bar{E}_{+0}$, so we have that

\begin{align}
        \bar{E}_{+0_{lin}} & = \frac{m\xi}{\sqrt{2}}, & \bar{E}_{+0_{circ}} = m\xi.
\end{align}
Similarly, the energy conservation law $\bar{E} = n\omega$ (see Eq.(\ref{eqn:ECL})) provides that the threshold photon number is $n_{0} = 2\bar{E}_0/\omega$. In the $\xi \gg 1$ regime, the ratios of the threshold and expected values are then
\begin{align}
        \frac{\langle n \rangle}{n_{0}}_{lin} = \frac{\langle\bar{E}_+\rangle}{\bar{E}_{+0}}_{lin}
        & = \frac{\xi}{2}, \label{eqn:RETE}\\
        \frac{\langle n \rangle}{n_{0}}_{circ} = \frac{\langle\bar{E}_+\rangle}{\bar{E}_{+0}}_{circ}
        & = \xi \sqrt{2}. \label{eqn:RNTE}
\end{align}
The result for the circular polarization is in agreement with previous results derived without use of the vicinal approximation \cite{AAMS03,MVG03}, and we have now shown that a similar result holds for the linear polarization case.

The positron energy spectra and photon absorption spectra are broad, due to the $1/\F\xi^4$ and $\omega^2/\F m^2 \xi^4$ factors in their respective exponents. However, compared to the typical scales of the spectra, defined by the expected values $\langle \bar{E}_+ \rangle$ and $\langle n \rangle$, the spectra are actually quite narrow in the tunneling regime. That is, from Eqs. (\ref{eqn:LPPES}) and  (\ref{eqn:CPPES}), the width of the energy spectra is
\begin{equation}
        \delta\bigg(\frac{\bar{E}_+}{m}\bigg) \sim \sqrt{F}\xi^2 \ll \xi^2 \sim \frac{\langle \bar{E}_+ \rangle}{m},
\end{equation}
and similarly from (\ref{eqn:LPPAS}) and (\ref{eqn:CPPAS}), the width of the photon absorption spectra is
\begin{equation}
        \delta(n)  \sim \sqrt{\F}\frac{m}{\omega}\xi^2 \ll \frac{m}{\omega}\xi^2 \sim \langle n \rangle.
\end{equation}
By Eqs. (\ref{eqn:RETE}) and (\ref{eqn:RNTE}), in the $\xi \gg 1 $ regime the threshold energy $\bar{E}_{+0} \ll \langle \bar{E}_+\rangle$ and threshold number $n_0 \ll \langle n \rangle$. The narrow width of the spectra in comparison to the their typical scales means that the spectra are strongly suppressed at the threshold. In other words, only leptons (absorbed photons) of energies (number) a factor of $\xi$ above the threshold value contribute significantly to the pair creation rate. By analogy to the (tunneling) multiphoton ionisation case \cite{AK85}, this effect is called above-threshold pair creation. This is shown in Figs. \ref{fig:LE} and \ref{fig:PAS}. Note that in these plots we choose $\omega \simeq 1$keV, and $\F = 0.02,\ 0.03$ based on possible operational parameters of XFELs \cite{Rin01}.

\begin{figure}[t]
\setlength{\unitlength}{0.120450pt}
\begin{picture}(2040,1620)(0,0)
\footnotesize
\thicklines \path(370,249)(411,249)
\thicklines \path(1916,249)(1875,249)
\put(329,249){\makebox(0,0)[r]{ 0}}
\thicklines \path(370,483)(411,483)
\thicklines \path(1916,483)(1875,483)
\put(329,483){\makebox(0,0)[r]{ 0.2}}
\thicklines \path(370,718)(411,718)
\thicklines \path(1916,718)(1875,718)
\put(329,718){\makebox(0,0)[r]{ 0.4}}
\thicklines \path(370,952)(411,952)
\thicklines \path(1916,952)(1875,952)
\put(329,952){\makebox(0,0)[r]{ 0.6}}
\thicklines \path(370,1186)(411,1186)
\thicklines \path(1916,1186)(1875,1186)
\put(329,1186){\makebox(0,0)[r]{ 0.8}}
\thicklines \path(370,1421)(411,1421)
\thicklines \path(1916,1421)(1875,1421)
\put(329,1421){\makebox(0,0)[r]{ 1}}
\thicklines \path(370,249)(370,290)
\thicklines \path(370,1538)(370,1497)
\put(370,166){\makebox(0,0){ 0}}
\thicklines \path(679,249)(679,290)
\thicklines \path(679,1538)(679,1497)
\put(679,166){\makebox(0,0){ 20}}
\thicklines \path(988,249)(988,290)
\thicklines \path(988,1538)(988,1497)
\put(988,166){\makebox(0,0){ 40}}
\thicklines \path(1298,249)(1298,290)
\thicklines \path(1298,1538)(1298,1497)
\put(1298,166){\makebox(0,0){ 60}}
\thicklines \path(1607,249)(1607,290)
\thicklines \path(1607,1538)(1607,1497)
\put(1607,166){\makebox(0,0){ 80}}
\thicklines \path(1916,249)(1916,290)
\thicklines \path(1916,1538)(1916,1497)
\put(1916,166){\makebox(0,0){ 100}}
\thicklines \path(370,249)(1916,249)(1916,1538)(370,1538)(370,249)
\put(82,893){\makebox(0,0)[l]{\rotatebox[origin=c]{90}{$dW/d\bar{E}_+$}}}
\put(1143,42){\makebox(0,0){$\bar{E}_+/mc^2$}}
\put(419,366){\makebox(0,0)[l]{A}}
\put(550,366){\makebox(0,0)[l]{B}}
\put(138,1538){\makebox(0,0)[l]{(a)}}
\put(480,835){\vector(0,-1){563}}
\put(535,835){\vector(0,-1){563}}
\thinlines \path(370,249)(370,249)(378,249)(386,249)(393,249)(401,249)(409,249)(417,249)(424,249)(432,249)(440,249)(448,249)(455,249)(463,249)(471,249)(479,249)(487,249)(494,249)(502,249)(510,249)(518,249)(525,249)(533,249)(541,249)(549,249)(556,249)(564,249)(572,249)(580,249)(588,249)(595,249)(603,249)(611,249)(619,249)(626,249)(634,249)(642,249)(650,249)(657,249)(665,249)(673,249)(681,249)(689,249)(696,249)(704,250)(712,250)(720,251)(727,252)(735,254)(743,257)(751,261)
\thinlines \path(751,261)(758,268)(766,277)(774,289)(782,307)(790,330)(797,360)(805,398)(813,446)(821,504)(828,572)(836,650)(844,737)(852,832)(859,931)(867,1031)(875,1129)(883,1218)(891,1296)(898,1358)(906,1399)(914,1419)(922,1416)(929,1390)(937,1342)(945,1275)(953,1194)(960,1101)(968,1003)(976,903)(984,804)(992,712)(999,627)(1007,551)(1015,486)(1023,431)(1030,386)(1038,350)(1046,322)(1054,301)(1061,285)(1069,274)(1077,266)(1085,260)(1093,256)(1100,253)(1108,252)(1116,251)(1124,250)(1131,250)(1139,249)
\thinlines \path(1139,249)(1147,249)(1155,249)(1162,249)(1170,249)(1178,249)(1186,249)(1193,249)(1201,249)(1209,249)(1217,249)(1225,249)(1232,249)(1240,249)(1248,249)(1256,249)(1263,249)(1271,249)(1279,249)(1287,249)(1294,249)(1302,249)(1310,249)(1318,249)(1326,249)(1333,249)(1341,249)(1349,249)(1357,249)(1364,249)(1372,249)(1380,249)(1388,249)(1395,249)(1403,249)(1411,249)(1419,249)(1427,249)(1434,249)(1442,249)(1450,249)(1458,249)(1465,249)(1473,249)(1481,249)(1489,249)(1496,249)(1504,249)(1512,249)(1520,249)(1528,249)
\thinlines \path(1528,249)(1535,249)(1543,249)(1551,249)(1559,249)(1566,249)(1574,249)(1582,249)(1590,249)(1597,249)(1605,249)(1613,249)(1621,249)(1629,249)(1636,249)(1644,249)(1652,249)(1660,249)(1667,249)(1675,249)(1683,249)(1691,249)(1698,249)(1706,249)(1714,249)(1722,249)(1730,249)(1737,249)(1745,249)(1753,249)(1761,249)(1768,249)(1776,249)(1784,249)(1792,249)(1799,249)(1807,249)(1815,249)(1823,249)(1831,249)(1838,249)(1846,249)(1854,249)(1862,249)(1869,249)(1877,249)(1885,249)(1893,249)(1900,249)(1908,249)(1916,249)
\thinlines \drawline[-50](370,249)(370,249)(378,249)(386,249)(393,249)(401,249)(409,249)(417,249)(424,249)(432,249)(440,249)(448,249)(455,249)(463,249)(471,249)(479,249)(487,249)(494,249)(502,249)(510,249)(518,249)(525,249)(533,249)(541,249)(549,249)(556,249)(564,249)(572,249)(580,249)(588,249)(595,249)(603,249)(611,249)(619,249)(626,249)(634,249)(642,249)(650,249)(657,249)(665,249)(673,249)(681,249)(689,249)(696,249)(704,249)(712,249)(720,249)(727,249)(735,249)(743,249)(751,249)
\thinlines \drawline[-50](751,249)(758,249)(766,249)(774,249)(782,249)(790,249)(797,249)(805,249)(813,249)(821,249)(828,249)(836,249)(844,249)(852,249)(859,249)(867,249)(875,249)(883,249)(891,249)(898,249)(906,249)(914,249)(922,249)(929,249)(937,249)(945,249)(953,249)(960,249)(968,249)(976,249)(984,249)(992,249)(999,249)(1007,250)(1015,250)(1023,250)(1030,250)(1038,250)(1046,250)(1054,251)(1061,251)(1069,252)(1077,252)(1085,253)(1093,253)(1100,254)(1108,255)(1116,256)(1124,257)(1131,259)(1139,260)
\thinlines \drawline[-50](1139,260)(1147,262)(1155,265)(1162,267)(1170,270)(1178,273)(1186,277)(1193,281)(1201,286)(1209,291)(1217,297)(1225,303)(1232,311)(1240,319)(1248,328)(1256,337)(1263,348)(1271,360)(1279,373)(1287,387)(1294,402)(1302,419)(1310,437)(1318,456)(1326,476)(1333,498)(1341,521)(1349,546)(1357,572)(1364,599)(1372,627)(1380,657)(1388,688)(1395,720)(1403,753)(1411,787)(1419,822)(1427,858)(1434,894)(1442,930)(1450,966)(1458,1003)(1465,1039)(1473,1075)(1481,1110)(1489,1144)(1496,1177)(1504,1209)(1512,1239)(1520,1268)(1528,1295)
\thinlines \drawline[-50](1528,1295)(1535,1319)(1543,1341)(1551,1361)(1559,1378)(1566,1393)(1574,1404)(1582,1413)(1590,1418)(1597,1421)(1605,1420)(1613,1416)(1621,1410)(1629,1400)(1636,1387)(1644,1372)(1652,1354)(1660,1333)(1667,1310)(1675,1284)(1683,1257)(1691,1228)(1698,1197)(1706,1164)(1714,1131)(1722,1096)(1730,1061)(1737,1025)(1745,989)(1753,952)(1761,916)(1768,880)(1776,844)(1784,809)(1792,774)(1799,740)(1807,708)(1815,676)(1823,645)(1831,616)(1838,588)(1846,561)(1854,536)(1862,512)(1869,489)(1877,468)(1885,448)(1893,429)(1900,412)(1908,396)(1916,381)
\thicklines \path(370,249)(1916,249)(1916,1538)(370,1538)(370,249)
\end{picture}
\setlength{\unitlength}{0.120450pt}
\begin{picture}(2040,1620)(0,0)
\footnotesize
\thicklines \path(370,249)(411,249)
\thicklines \path(1916,249)(1875,249)
\put(329,249){\makebox(0,0)[r]{ 0}}
\thicklines \path(370,483)(411,483)
\thicklines \path(1916,483)(1875,483)
\put(329,483){\makebox(0,0)[r]{ 0.2}}
\thicklines \path(370,718)(411,718)
\thicklines \path(1916,718)(1875,718)
\put(329,718){\makebox(0,0)[r]{ 0.4}}
\thicklines \path(370,952)(411,952)
\thicklines \path(1916,952)(1875,952)
\put(329,952){\makebox(0,0)[r]{ 0.6}}
\thicklines \path(370,1186)(411,1186)
\thicklines \path(1916,1186)(1875,1186)
\put(329,1186){\makebox(0,0)[r]{ 0.8}}
\thicklines \path(370,1421)(411,1421)
\thicklines \path(1916,1421)(1875,1421)
\put(329,1421){\makebox(0,0)[r]{ 1}}
\thicklines \path(370,249)(370,290)
\thicklines \path(370,1538)(370,1497)
\put(370,166){\makebox(0,0){ 0}}
\thicklines \path(563,249)(563,290)
\thicklines \path(563,1538)(563,1497)
\put(563,166){\makebox(0,0){ 50}}
\thicklines \path(757,249)(757,290)
\thicklines \path(757,1538)(757,1497)
\put(757,166){\makebox(0,0){ 100}}
\thicklines \path(950,249)(950,290)
\thicklines \path(950,1538)(950,1497)
\put(950,166){\makebox(0,0){ 150}}
\thicklines \path(1143,249)(1143,290)
\thicklines \path(1143,1538)(1143,1497)
\put(1143,166){\makebox(0,0){ 200}}
\thicklines \path(1336,249)(1336,290)
\thicklines \path(1336,1538)(1336,1497)
\put(1336,166){\makebox(0,0){ 250}}
\thicklines \path(1530,249)(1530,290)
\thicklines \path(1530,1538)(1530,1497)
\put(1530,166){\makebox(0,0){ 300}}
\thicklines \path(1723,249)(1723,290)
\thicklines \path(1723,1538)(1723,1497)
\put(1723,166){\makebox(0,0){ 350}}
\thicklines \path(1916,249)(1916,290)
\thicklines \path(1916,1538)(1916,1497)
\put(1916,166){\makebox(0,0){ 400}}
\thicklines \path(370,249)(1916,249)(1916,1538)(370,1538)(370,249)
\put(82,893){\makebox(0,0)[l]{\rotatebox[origin=c]{90}{$dW/d\bar{E}_+$}}}
\put(1143,42){\makebox(0,0){$\bar{E}_+/mc^2$}}
\put(444,366){\makebox(0,0)[l]{A,\ B}}
\put(138,1538){\makebox(0,0)[l]{(b)}}
\put(409,835){\vector(0,-1){563}}
\put(428,835){\vector(0,-1){563}}
\thinlines \path(370,249)(370,249)(376,249)(383,249)(389,249)(396,249)(402,249)(409,249)(415,249)(422,249)(428,249)(435,249)(441,249)(448,249)(454,249)(461,249)(467,249)(473,249)(480,249)(486,249)(493,249)(499,249)(506,249)(512,249)(519,249)(525,249)(532,249)(538,249)(545,249)(551,249)(558,249)(564,249)(571,249)(577,249)(583,249)(590,249)(596,249)(603,249)(609,249)(616,249)(622,249)(629,249)(635,249)(642,249)(648,249)(655,249)(661,249)(668,249)(674,249)(680,249)(687,249)
\thinlines \path(687,249)(693,249)(700,249)(706,250)(713,250)(719,251)(726,252)(732,253)(739,255)(745,258)(752,262)(758,267)(765,275)(771,284)(778,297)(784,313)(790,333)(797,358)(803,389)(810,427)(816,471)(823,522)(829,580)(836,646)(842,718)(849,795)(855,876)(862,959)(868,1043)(875,1123)(881,1199)(887,1267)(894,1325)(900,1371)(907,1403)(913,1419)(920,1419)(926,1403)(933,1371)(939,1325)(946,1267)(952,1199)(959,1123)(965,1042)(972,959)(978,876)(985,795)(991,718)(997,646)(1004,580)(1010,522)
\thinlines \path(1010,522)(1017,471)(1023,427)(1030,389)(1036,358)(1043,333)(1049,313)(1056,297)(1062,284)(1069,274)(1075,267)(1082,262)(1088,258)(1094,255)(1101,253)(1107,252)(1114,251)(1120,250)(1127,250)(1133,249)(1140,249)(1146,249)(1153,249)(1159,249)(1166,249)(1172,249)(1179,249)(1185,249)(1192,249)(1198,249)(1204,249)(1211,249)(1217,249)(1224,249)(1230,249)(1237,249)(1243,249)(1250,249)(1256,249)(1263,249)(1269,249)(1276,249)(1282,249)(1289,249)(1295,249)(1301,249)(1308,249)(1314,249)(1321,249)(1327,249)(1334,249)
\thinlines \path(1334,249)(1340,249)(1347,249)(1353,249)(1360,249)(1366,249)(1373,249)(1379,249)(1386,249)(1392,249)(1399,249)(1405,249)(1411,249)(1418,249)(1424,249)(1431,249)(1437,249)(1444,249)(1450,249)(1457,249)(1463,249)(1470,249)(1476,249)(1483,249)(1489,249)(1496,249)(1502,249)(1508,249)(1515,249)(1521,249)(1528,249)(1534,249)(1541,249)(1547,249)(1554,249)(1560,249)(1567,249)(1573,249)(1580,249)(1586,249)(1593,249)(1599,249)(1606,249)(1612,249)(1618,249)(1625,249)(1631,249)(1638,249)(1644,249)(1651,249)(1657,249)
\thinlines \path(1657,249)(1664,249)(1670,249)(1677,249)(1683,249)(1690,249)(1696,249)(1703,249)(1709,249)(1715,249)(1722,249)(1728,249)(1735,249)(1741,249)(1748,249)(1754,249)(1761,249)(1767,249)(1774,249)(1780,249)(1787,249)(1793,249)(1800,249)(1806,249)(1813,249)(1819,249)(1825,249)(1832,249)(1838,249)(1845,249)(1851,249)(1858,249)(1864,249)(1871,249)(1877,249)(1884,249)(1890,249)(1897,249)(1903,249)(1910,249)(1916,249)
\thinlines \drawline[-50](370,249)(370,249)(376,249)(383,249)(389,249)(396,249)(402,249)(409,249)(415,249)(422,249)(428,249)(435,249)(441,249)(448,249)(454,249)(461,249)(467,249)(473,249)(480,249)(486,249)(493,249)(499,249)(506,249)(512,249)(519,249)(525,249)(532,249)(538,249)(545,249)(551,249)(558,249)(564,249)(571,249)(577,249)(583,249)(590,249)(596,249)(603,249)(609,249)(616,249)(622,249)(629,249)(635,249)(642,249)(648,249)(655,249)(661,249)(668,249)(674,249)(680,249)(687,249)
\thinlines \drawline[-50](687,249)(693,249)(700,249)(706,249)(713,249)(719,249)(726,249)(732,249)(739,249)(745,249)(752,249)(758,249)(765,249)(771,249)(778,249)(784,249)(790,249)(797,249)(803,249)(810,249)(816,249)(823,249)(829,249)(836,249)(842,249)(849,249)(855,249)(862,249)(868,249)(875,249)(881,249)(887,249)(894,249)(900,249)(907,249)(913,249)(920,249)(926,249)(933,249)(939,249)(946,249)(952,249)(959,249)(965,249)(972,249)(978,249)(985,249)(991,249)(997,249)(1004,250)(1010,250)
\thinlines \drawline[-50](1010,250)(1017,250)(1023,250)(1030,250)(1036,250)(1043,250)(1049,251)(1056,251)(1062,251)(1069,251)(1075,252)(1082,252)(1088,253)(1094,253)(1101,254)(1107,255)(1114,256)(1120,257)(1127,258)(1133,259)(1140,261)(1146,262)(1153,264)(1159,266)(1166,268)(1172,271)(1179,273)(1185,276)(1192,280)(1198,284)(1204,288)(1211,292)(1217,297)(1224,303)(1230,309)(1237,315)(1243,322)(1250,330)(1256,338)(1263,347)(1269,357)(1276,367)(1282,379)(1289,391)(1295,403)(1301,417)(1308,432)(1314,447)(1321,464)(1327,481)(1334,499)
\thinlines \drawline[-50](1334,499)(1340,519)(1347,539)(1353,560)(1360,582)(1366,605)(1373,629)(1379,654)(1386,679)(1392,706)(1399,733)(1405,761)(1411,789)(1418,818)(1424,848)(1431,878)(1437,908)(1444,938)(1450,968)(1457,999)(1463,1029)(1470,1059)(1476,1088)(1483,1117)(1489,1146)(1496,1173)(1502,1200)(1508,1226)(1515,1250)(1521,1274)(1528,1296)(1534,1316)(1541,1335)(1547,1352)(1554,1368)(1560,1381)(1567,1393)(1573,1403)(1580,1410)(1586,1416)(1593,1419)(1599,1421)(1606,1420)(1612,1417)(1618,1412)(1625,1405)(1631,1396)(1638,1384)(1644,1371)(1651,1356)(1657,1340)
\thinlines \drawline[-50](1657,1340)(1664,1321)(1670,1301)(1677,1279)(1683,1256)(1690,1232)(1696,1206)(1703,1180)(1709,1153)(1715,1124)(1722,1095)(1728,1066)(1735,1036)(1741,1006)(1748,976)(1754,945)(1761,915)(1767,885)(1774,855)(1780,826)(1787,796)(1793,768)(1800,740)(1806,713)(1813,686)(1819,660)(1825,635)(1832,611)(1838,588)(1845,565)(1851,544)(1858,524)(1864,504)(1871,485)(1877,468)(1884,451)(1890,435)(1897,421)(1903,407)(1910,394)(1916,381)
\thicklines \path(370,249)(1916,249)(1916,1538)(370,1538)(370,249)
\end{picture}
        \caption{Positron energy spectra (normalized to unity at their maxima) for (a) the case of linear polarization, and (b) the case of circular polarization. Here $\omega = 0.002m$ with $\F = 0.02$ (solid line) or $\F = 0.03$ (dashed line), so that $\xi =  10$ or $15$ respectively. Arrow A (B) indicates the location of the threshold energy $\bar{E}_{+0}$ for $\F = 0.02\ (0.03)$. }
        \label{fig:LE}
\end{figure}
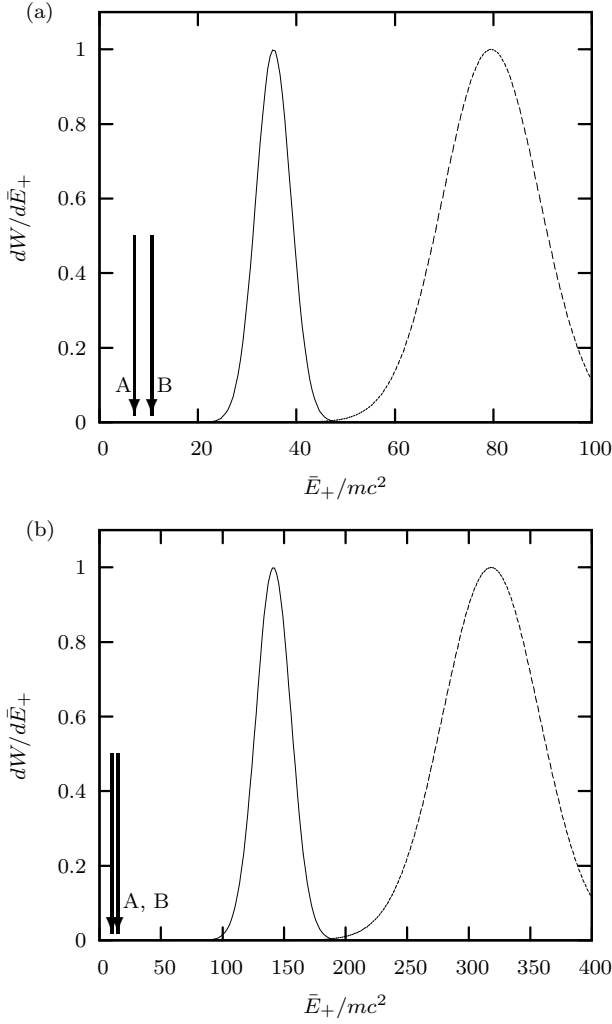

\begin{figure}[t]
        \input{LPAS.tex}
        \input{CPAS.tex}
        \caption{Photon absorption spectra (normalized to unity at their maxima) for (a) the case of linear polarization, and (b) the case of circular polarization. Here $\omega = 0.002m$ with $\F = 0.02$ (solid line) or $\xi = 0.03$ (dashed line), so that $\xi =  10$ or $15$ respectively. Arrow A (B) indicates the location of the threshold number $n_{0}$ at $\F = 0.02\ (0.03)$.}
        \label{fig:PAS}
\end{figure}

\subsection{Angular Distributions}

\begin{figure}[t]
        \input{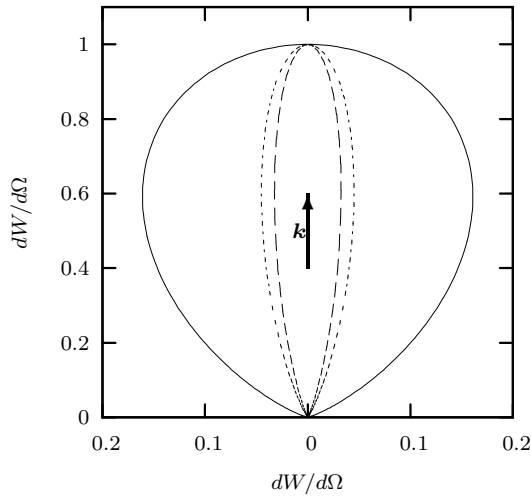}
        \caption{Cross-sectional polar plots of the angular distribution (normalized to unity at its maximum) for the linear polarization case for various fixed azimuthal angles $\varphi$. The polar coordinates are $(dW/d\Omega, \theta)$, where the polar angle $\theta$ is measured from the direction of the light wavevector $\bm{k}$, indicated by the arrow. Plots are shown for: $\bm{k}$ - $\E$ or $\varphi = 0$ plane (solid line); $\bm{k}$ - $\B$ or $\varphi = \pi/2$ plane (dashed line); and $\varphi = \pi/4$ plane (dotted line). Here $\omega = 0.002m$ and $\F = 0.01$ so that $\xi = 5$ and $\G = 0.25$. The distribution has a sharp peak at $\theta = 0$, and is suppressed in the magnetic field direction.}
        \label{fig:LAS}
\end{figure}

\begin{figure}[t]
        \input{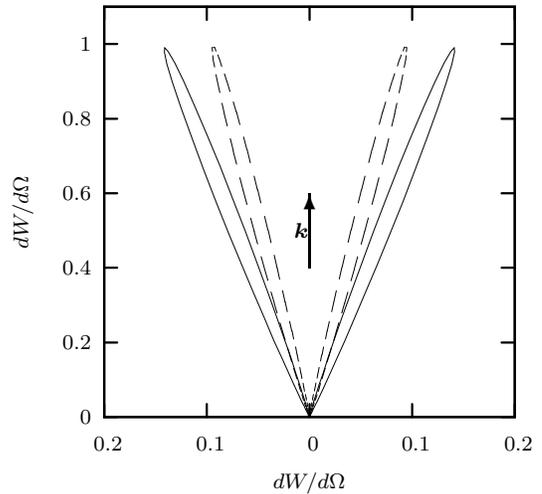}
        \caption{Cross-sectional polar plots of the angular distribution (normalized to unity at its maximum) for the circular polarization at arbitrary fixed azimuthal angle $\varphi$. The polar coordinates are $(dW/d\Omega, \theta)$, where the polar angle $\theta$ is measured from the direction of the light wavevector $\bm{k}$, indicated by the arrow. Here $\omega = 0.002m$ and $\F = 0.02$ (solid line) or $\F = 0.03$ (dashed line) so that $\xi = 10$ or $15$ respectively. The plots clearly exhibit very sharp peaks. Due to cylindrical symmetry the full distribution is obtained by rotating about $\bm{k}$ to produce a thin-walled cone.}
        \label{fig:CAS}
\end{figure}

The angular distribution (\ref{eqn:LPARS}) in the linear polarisation case is a Gaussian function of the polar angle $\theta$, so that the expected direction of positron emission is along the $\theta = 0$ direction. Moreover, the $\xi^2$ factor of the $\sin^2\varphi$ term in the exponent means that the the angular distribution is strongly suppressed in the direction of the magnetic field ($\varphi = \pi/2$). In other words, the spectrum is flattened into the electric field plane (Fig. \ref{fig:LAS}).

Eq.(\ref{eqn:LPARS}) features a factor $1/\G$ in its exponent. Consequently, it is applicable literally only in the special case $\G \ll 1$ of the tunneling regime. In this case the spectrum has a sharp peak at $\theta = 0$ (Fig. \ref{fig:LAS}), so that the positron and electrons are overwhelmingly emitted in the same direction. We expect though that the spectrum should exhibit similar main features within the general tunneling regime (arbitrary $\G$, $\F \ll 1$): a peak in the $\theta = 0$ direction and flattening into the electric field plane.

The expression for the angular distribution in the circular polarization case (\ref{eqn:CPARS}) gives the mean polar angle of positron emission  $\theta = \sqrt{2}/\xi$. Due to the $\xi^2/\F$ factor in the exponent, this distribution is very sharp. That is the width of the distribution is
\begin{equation}
        \delta\theta \sim \frac{\sqrt{\F}}{\xi} \lll 1~.
\end{equation}
Hence this distribution, which is symmetric in azimuthal angle $\varphi$, is a thin-walled cone with axis of symmetry being the light wavevector and with conical angle $\sqrt{2}/\xi$ (Fig. \ref{fig:CAS}). Note also that $\varphi_- = \varphi_+ - \pi$ at the momentum saddle point (see Eq. \ref{eqn:CPASPC}),  so the electron and positron are emitted in antipodal directions along this cone.

\section{Conclusion}
We considered the electron-positron pair production by Coulomb and laser fields in the tunneling regime describing the process in simple analytical terms, and deriving expressions for angular and energy distributions. The energy of the pair proves to be high, well above the threshold for the pair creation for any polarization. The angular distributions strongly depend on the polarization. For the linear case both leptons move predominantly along the direction of the laser beam.
For the circular polarization the pair is distributed in a thin-walled cone with axis of symmetry along the direction of the light beam, and with small conical angle. The leptons follow antipodal directions on the cone. The vicinal approximation suggested in this work proves to be very convenient for analytical calculations.  

\acknowledgments
The support of the Australian Research Council is acknowledged.  

\appendix

\section{Spin Factors}
\label{app:SF}
In all polarization cases, the differential pair creation rate $dW$ includes a spin factor
\begin{equation}
        K = \sum_{s_{\pm}}|\mathcal{K}(t_0)|^2~,
\end{equation}
where $\mathcal{K}(t)$ is defined in Eq. (\ref{eqn:SFD}), and the subscript `$0$' indicates evaluation at the appropriate saddle points in the time and momenta variables.

In the linear polarization case, from Eq. (\ref{eqn:QMVA}) one finds that at the saddle points $t_1$ and $P_0$ in the $\xi \gg 1$ regime
\begin{equation}
        Q^*_+ = \bar{Q}_- = 1 + i\frac{\sqrt{3}}{2}(\gamma^0 - \gamma^3)\gamma^1 \equiv Q_0.
\end{equation}
It is straightforward to show that the matrix $Q_0$ has the property $Q_0 = \bar{Q}_0$. The spin factor is then
\begin{equation}
        \label{eqn:SFGE}
        K = \frac{1}{4\varepsilon_0^2}\sum_{s_{\pm}}|\bar{u}_{\bm{p}_-}Q_0\gamma^0Q_0u_{-\bm{p}_+}|^2,
\end{equation}
where at the saddle points $t_1$ and $P_0$, the lepton energy $\varepsilon_{0} = 3m/2/\sqrt{2}$ by Eq. (\ref{eqn:PZE}). However, the free Dirac spinors $u_{\pm\bm{p}_{\mp}}$ satisfy $(\gamma p_{0\pm})u_{\pm\bm{p}_{\mp}} =  m u_{\pm\bm{p}_{\mp}}$ (see Eq. \ref{eqn:DE})). Hence we may write the spin factor as a trace of projection operators
\begin{align}
        K
        & = \frac{8}{9}\frac{1}{4m^2}\mbox{Tr}\bigg[\Big((\gamma p_{0-}) - m\Big)Q_0\gamma^0Q_0\notag\\
        & \quad \quad \times \Big((\gamma p_{0+}) + m\Big)Q_0\gamma^0Q_0\bigg] = \frac{8}{9}.
\end{align}
In the circular polarization case, once more from Eq. (\ref{eqn:QMVA}) one finds that at the saddle points $t_0$ and $P_0$ in the $\xi \gg 1$ regime
\begin{equation}
        Q^{*}_+ = \bar{Q}_- = 1 + \frac{1}{\sqrt{2}}\left(\gamma^0 -\gamma^3\right)
        \bigg(\xi\gamma^1 + i\gamma^2\sqrt{\frac{3}{2}}\bigg) \equiv Q_0.
\end{equation}
This time, $Q_0 \not= \bar{Q}_0$, and the lepton energy $\varepsilon_0 = m\xi^2/\sqrt{2}$. It follows then from Eq. (\ref{eqn:SFGE}) by similar reasoning to the linear polarisation case that
\begin{align}
        K
        & = \frac{2}{\xi^4}\frac{1}{4m^2}\mbox{Tr}\bigg[\Big((\gamma p_{0-}) - m\Big)Q_0\gamma^0Q_0\notag\\
        & \quad \quad \times \Big((\gamma p_{0+}) + m\Big)\bar{Q}_0\gamma^0\bar{Q}_0\bigg] = \frac{2}{\xi^4}.
\end{align}

\bibliography{pairs}

\end{document}